\documentclass{aa}
\usepackage[utf8]{inputenc}

\usepackage{graphicx}
\usepackage{newtxtext}
\usepackage{newtxmath}

\usepackage{multirow}
\usepackage{longtable}
\usepackage{array}
\usepackage{lipsum}
 \usepackage{ulem} 
\usepackage[usenames,dvipsnames,svgnames,table]{xcolor} 

\usepackage{hyperref}
\hypersetup{
  colorlinks=true,
  citecolor=blue,
  linkbordercolor={1 0 0},
  linkcolor=blue,
  urlcolor=blue,
  breaklinks=true
}

\newcommand{\rrlyr}{RR~Lyrae\ }

\newcommand{\rrly}{RR~Lyr\ }
\newcommand{\rrlyn}{RR~Lyr}
\newcommand{\ha}{H$\alpha$\ }
\newcommand{\han}{H$\alpha$}
\newcommand{\hb}{H$\beta$\ }
\newcommand{\hg}{H$\gamma$\ }
\newcommand{\hgn}{H$\gamma$}

\newcommand{\kms}{\;km\,s$^{-1}$\ }
\newcommand{\kmsn}{\;km\,s$^{-1}$}

\renewcommand{\aap}{A\textit{\&}A} 

\begin{document} 

\title{Atmospheric dynamics and shock waves in \rrly}

\subtitle{I. The three \ha emissions}

\author{D. Gillet  \inst{1}
\and    F.L. Sefyani   \inst{2}
\and    A. Benhida \inst{2}
\and    B. Mauclaire\inst{3}
\and    D. El Youssoufi \inst{4}
\and    T.~de~France\inst{5}
\and    Z. Benkhaldoun \inst{2}
\and    P. Lailly \inst{5}
}
\institute{Observatoire de Haute-Provence - CNRS/PYTHEAS/Universit\'e d'Aix-Marseille, 04870 Saint Michel l'Observatoire, France\thanks{Based in part on observations made at the Observatoire de Haute Provence (CNRS), France and on observations obtained at the Ouka\"imeden Observatory in the High Atlas mountains, 78\,km south of Marrakech and operated by the Facult\'e des Sciences Semlalia, D\'epartement de Physique, LPHEA, Marrakech, Morocco.}
\email{denis.gillet@osupytheas.fr}
\and       Observatoire d'Ouka\"imeden, Facult\'e des Sciences et Techniques, D\'epartement de Physique, LPHEA, Marrakech, Morocco
\email{sefyani@uca.ma}
\and Observatoire du Val de l'Arc, route de Peynier, 13530 Trets, France \email{bma.ova@gmail.com}
\and Leibniz-Institut für Astrophysik Potsdam (AIP), An der Sternwarte 16, D-14482 Potsdam, Germany
\and Observatoire des Tourterelles, 34140 Mèze, France \
\and Observatoire Canigou, route de Pau, 64445 Pau, France
}

\date{Received July, 2023; accepted xxxx}

\abstract
{Although spectroscopic observations of RR\,Lyrae stars have been underway for almost a century, the fact that the hydrogen line exhibits three successive emissions in each pulsation cycle is still a very recent discovery.}
{The purpose of the present study is to clarify the physical origin of these three emissions and their connection to atmospheric dynamics and to examine the influence of Blazhko modulation on their intensity.}
{We used 2437 high-resolution spectra over a total of 81 nights taken by the  ELODIE spectrograph (Haute Provence Observatory, France) in the years 1994-1997, rounded out with a 2015 run from  Ouka\"{i}meden Observatory (Morocco). We performed a detailed analysis of the line profile variations over the whole pulsation cycle. 
}
{Based on the blueshift of the main \ha emission, the velocity of the hypersonic shock front was estimated at between 100 and 150 $\pm 10$ \kmsn (Mach number between 10 and 15).
It has been established that the shock velocity increases from the
minimum Blazhko to its maximum and afterward, it gradually decreases to the Blazhko minimum to start growing again.
This observational result is consistent with the shock model proposed in 2013 to explain the Blazhko effect.
The intensity of the \ha emission increases with the shock velocity up to a maximum value around 137\kms and then decreases as the shock velocity increases further. This effect would be the consequence of the increasingly important ionization of the atoms in the radiative shock wake.
The second (blueshifted) \ha emission is the consequence of an approximately constant supersonic compression (Mach number between 2 and 3) of the upper atmosphere falling onto the photospheric layers, during 3 to 16\% of the pulsation period.
Finally, the third \ha emission (P-Cygni profile) would be the consequence of the expansion of the high atmosphere induced by the shock wave during its final weakening.
}
{}


\maketitle

\section{Introduction}

The radially pulsating star RR\,Lyr is one of the best known and studied objects of its kind. Its light curve of RR\,Lyr exhibits a (variable) period around 0.5667 days or, equivalently, 13.6 h (the pulsation period), but it also presents light modulations
with a (variable) period around 39 days. These amplitude
and phase modulations are widely known as the Blazhko effect \citep{Bla1907}. In this paper, the pulsation phase is noted as $\varphi,$ where $\varphi=0.0$ corresponds to the 
luminosity maximum. The Blazhko phase is noted $\psi$ and has its maximum at $\psi=0.0$ when the highest luminosity amplitude is observed during a Blazhko cycle.

At present, it has been established that strong shock waves cross the atmosphere of RR Lyrae stars during the periodic movements of pulsation. The observational facts in favor of these waves include: the doubling of a few absorption metallic lines, the Van Hoof phenomenon \citep{VHS1953}, and, certainly the most obvious, the appearance of hydrogen lines in emission, especially \han. 

It has now been clearly established that during each
cycle of pulsation, the \ha line  presents  three components in emission.
The first, known as the "first apparition," is the most intense of the three. This blueshifted emission was first reported by \cite{Str1948}.  It occurs immediately before the luminosity maximum, around the pulsation phase of $\varphi=0.92$, and may be caused by a strong shock wave, hereafter called "the main shock" that is induced by the $\kappa$-mechanism. A series of nonlinear nonadiabatic pulsation models of RR Lyrae that allow for the propagation of shock waves in its atmosphere have been calculated by \citet{FG1997}. To date, this study has provided the most detailed analysis of the physical origins of the different shocks occurring during a pulsation cycle. The self-consistent calculation of \ha emission line profile within a radiative shock wake propagating in a stellar atmosphere was carried out by \citet{FG2004} and it offers a suitable estimate of the possible shock velocity range.

The second \ha emission, namely, the "second apparition" was discovered by \citet{GC1988} as a small blueshifted emission component. It appears during the luminosity bump, that is, just before the luminosity minimum near the pulsation phase of $\varphi=0.72$. \citet{GC1988} attributed this second emission to the inward propagation of a shock wave produced by the strong compression of high atmospheric layers on the deepest ones during the end of the ballistic movement of the atmosphere.

Finally, the third \ha emission, that is,  the "the third apparition" was first highlighted in a dozen RR\,Lyrae stars by \citet{Pre2011} using a large telescope. This weak and redshifted emission appears within the \ha absorption profile between the pulsation phase $\varphi=0.20$ and $\varphi=0.40,$ namely, when the atmosphere approaches its maximum expansion.
\citet{Gillet2017} proposed that the physical origin of this emission is probably due to the infalling motion of the highest atmospheric layers, which compresses and heats the gas that is located immediately above the rising shock wave.

The pulsation of RR\,Lyr essentially concerns  the photosphere and the atmospheric layers, which represent only a small percent of the mass of the star. \citet{Gillet2019} decomposed the variation of atmospheric dynamics during a pulsation cycle into six main phases. First, the emergence of the main shock from the photosphere occurs around $\varphi=0.89,$ which explains why a blueshifted \ha emission component suddenly appears within the \ha profile. Then, between $\varphi=0.89$ and $\varphi=0.93$, this \ha emission is visible, indicating that the shock is in a radiative phase. Near $\varphi=0.32$, the sodium layer reaches its maximum expansion while the highest part of the atmosphere where \ha is formed is still ascending. The latter reaches its maximum expansion around $\varphi=0.46$. Consequently, between $\varphi=0.32$ and $\varphi=0.46$, a two-step infalling motion takes place within the atmosphere. Finally, a strong phostospheric compression is observed around $\varphi=0.36,$  thus forming the "third apparition." 

One of the objectives of this work is to check if the variations and in particular the appearance of the three successive \ha emission components is compatible with this scheme of atmospheric dynamics. Moreover, we consider the influence of the Blazhko effect on the intensity of the shock waves and on the intensity of the three \ha emissions occurring during a pulsation cycle. The Blazhko Phenomenon was first observed by Sergey Blazhko in 1907 in the RR Lyrae type variable star RW Draconis \citep{Bla1907} and it is a long-period modulation in period and amplitude of the light curve. Today, the physical origin behind the Blazhko effect is currently still a matter of debate.

\medskip

We present hereafter a large set of observations of profile variations of the \ha line in \rrly itself, including the three \ha emissions. 
In Sect.\,2, we describe the observed data. Their reduction processes and analysis is presented in Sect.\,3,
in particular, the occurrence of the three \ha emissions. The consequences of the line profile variations of the three \ha emissions are discussed in Sect.\,4, especially with respect to their effects on the dynamical structure of the atmosphere.
Finally, we draw our conclusions in Sect.\,5.

\section{Observations and data analysis}

The spectral data analyzed in this work are based on two distinct and independent sets of spectroscopic observations. 
The first consists of 1268 spectra recorded over 40 nights of observation in 1994-1997 and the second of 1169 spectra over 41 nights in 2013 and 2015. The first group of data 
is better respectively in terms of both the mean signal-to-noise ratio (S/N) of $\thicksim 30$ and resolving power 42,000 instead of S/N of $\thicksim 10$ and 10,700 around 5897\,\AA\ for the second. 
 The S/N per pixel, subsequently denoted S/N, is estimated with $2\sigma$ around 6557\,\AA\ over 2.5\,\AA\ by computing the ratio mean flux over standard deviation.
All these data that spread over 21 years come from 81 different cycles of pulsation, with the knowledge that no cycle was observed in full 
since the pulsation period RR Lyr (13.6\,h) is always greater than the length of any night.
The 1994-1997 data are spread over ten different Blazhko cycles ($\thicksim 39$\,d), while all those from 2015 come from about the same cycle if we do not take into account the 2013 observation made on a single night.
The residual flux(\%) for the continuum of emission lines (H$\alpha$, He\,I) is summarized in Table \ref{longobslog}.

\subsection{Data acquisition}

The campaign took place in 2015 at the Ouka\"{i}meden Observatory, on the High Atlas Mountains, at an altitude of 2700\,m. The instruments set up includes 
two telescopes pickabacked. One 0.35\,m SCT dedicated to spectroscopy along with a fiber echelle spectrograph from Shelyak \citep{TC2010} reaching a resolving power $R\thicksim 10,700$ around 5897\,\AA\ over a spectral domain from 4100 to 7200\,\AA. The S/N ranges from 5 to 26 for exposure time between 300\,s and 400\,s, giving a temporal resolution  $\Delta\,t/P = 0.6$\%.
For the photometry, a 0.25\,m SCT with a CCD camera were used with various color filters (U-B-V). 
The advantage of this setup is obvious by the fact that spectroscopy and photometry were carried out simultaneously. Photometry was mainly used 
for the determination of pulsation maxima and thus the calculation of the pulsation phases.

RR Lyr observations from August\,3, 1994 and August\,10, 1997, recorded with the \textsc{Elodie} spectrograph \citep{Baranne1996}, were re-analyzed to achieve comparison. The resolving power is close to 42,000 on the whole visible domain (3900-6800\,\AA) with a signal-to-noise between 11 and 55. 
\textsc{Elodie} is connected by optical fiber to the Cassegrain telescope of 1.93\,m diameter. Each exposure had a duration between 7 and 10\,mn, giving a temporal resolution of $\Delta\,t/P$ = $1\pm 0.1$\%. 


{\setlength{\tabcolsep}{4.5pt}
\begin{table*}
\centering
\caption{Observations log of the RR\,Lyr spectra around the maximum luminosity. When the quality of the data allowed to determine the shock velocity, the nights are marked 
in bold.
Columns provide the date of the night, the observatory (Obs.), the number, $N,$ of observed spectra, the exposure time, T$_{\textrm{exp}}$, or the range of exposure time used during the night. The value in parenthesis is the exposure time corresponding to the spectrum of the maximum of H$\alpha$. 
Then, the corresponding Blazhko phase, $\psi$, of the night, the pulsation phase $\varphi_{emi}$ of the maximum of the emission of H$\alpha$ called "the first apparition", the estimated S/N per pixel with $2\sigma$ around 6557\AA\, over 2.5\AA, the maximum shock velocity (\kmsn) near $\varphi\simeq 0.9$ and the maximum intensity of the H$\alpha$ and He\,I D3 emission components with respect to the continuum. When the intensity number is in parenthesis, it means that the presence of the emission is doubtful in terms of the noise level or continuum; when it is equal to zero, no emission was observed. When $I$ is in square brackets, it means that the spectrum on which their measurements are made is either the first or the last of the night and therefore may not match at the maximum value.}
\label{longobslog}
\begin{tabular}{cccccccccc}
\hline \hline
Night & Obs. & N & T$_{\textrm{exp}}$ &  $\psi$ &   $\varphi_{emi}$ & S/N  & $V_{shock}$ & I(\%)  & I(\%) \\
{\scriptsize (yyyy-mm-dd)} &   &   & (sec)  & Blazhko & pulsation  & &   (\kms)
       & H$\alpha$ & HeI D3 \\
\hline 
\textbf{1994-08-03} & OHP  &    26      & 450-480 (480) &   0.44        &   0.956 & 55 & \textbf{111} &  \textbf{20}  & 4 \\
1994-08-04 & OHP  &     35      & 420-600 (480) &  0.47 &   0.931 & 36 & - & [18]  & [0]  \\
\textbf{1995-08-07} & OHP  &    18  & 300-480 (360) &   0.86    &  0.890        & 22 & \textbf{131} & -  & 0 \\
1995-08-08 & OHP  &     32      & 300-540 (540) &  0.89 &  0.877        & 37 & - & 4  & 0  \\
1995-09-05 & OHP  &     48      & 180-600 (360) &   1.60        &  0.915        & 27 & - & 7  & 6  \\
\textbf{1995-10-17} & OHP  &    38      &   300             &   2.67    &  0.890  & 32 & \textbf{128} & \textbf{22}  & 0  \\
\textbf{1996-06-24} & OHP  & 30 & 300-500 (500) &  0.12 &  0.882        & 34 & \textbf{122} & \textbf{27}  & (2) \\
\textbf{1996-06-25} & OHP  & 34 & 240-540 (420) &  0.15 &  0.873        & 48 & \textbf{119} &  \textbf{18}  & 0  \\
\textbf{1996-08-01} & OHP  & 56 & 180-360 (300) &  1.10 &  0.876        & 31 & \textbf{143} & \textbf{33}  & 3  \\
\textbf{1996-08-02} & OHP  &    20      & 420-600 (600) &  1.12 &  0.870        & 24 & \textbf{131} & \textbf{28}  & 3  \\
\textbf{1996-08-11} & OHP  & 36 & 420-600 (480) &  1.36 &  0.907        & 33 &  \textbf{120} & \textbf{29}  & 5  \\
\textbf{1997-05-30} & OHP  & 28 & 300-600 (420) &  0.95 &  0.877        & 11 & \textbf{135} & -  & 0  \\
\textbf{1997-06-11} & OHP  &    38      & 180-300 (180) &  1.26 &  0.887        & 24 & \textbf{129} & \textbf{33}  & 1  \\
\textbf{1997-08-05} & OHP  &    37      & 300-480 (300) &  2.66 &  0.896        & 39 & \textbf{139} & \textbf{40}  & 8  \\
\textbf{1997-08-06} & OHP  & 58 &   300             &  2.69     &  0.898        & 29 & \textbf{134} & \textbf{40}  & 8  \\
\textbf{1997-08-09} & OHP  & 48 & 300-480 (300) &  2.76 &  0.883        & 26 & \textbf{149} & \textbf{25}  & 4  \\
\textbf{1997-08-10} & OHP  & 38 & 300-600 (300) &  2.79 &  0.885        & 15 & \textbf{146} & \textbf{24}  & 0  \\
\textbf{1997-08-30} & OHP  & 65 &   300             &  3.30     &  0.884        & 36 & \textbf{124} &  \textbf{21}  & 5  \\
\textbf{1997-08-31} & OHP  & 60 & 180-300 (240) &  3.33 &  0.881        & 29 & \textbf{121} & \textbf{27}  & 0  \\
\textbf{2013-10-12} & Ouka &    40  &   300             &   0.16        &  0.922 & 13 & \textbf{117}  & \textbf{43}  & 7  \\
2015-10-29 & Ouka & 58  &   300             &   0.33    &  0.921        & 19 & -  & 34  & 3  \\
\textbf{2015-11-06} & Ouka & 23 & 300   &   0.53        &  0.917        & 26 & \textbf{119} & \textbf{31}  & 3  \\
2015-11-10 & Ouka & 43  &   300             &   0.63    &  0.919        & 17 & - & 32  & 7  \\
2015-11-11 & Ouka & 32  &   300             &   0.66    &  0.915        & 5 & - & 27  & 3  \\
2015-11-14 & Ouka &     48      &   300             &   0.74    &  0.920        & 6 & - & 44  & (2) \\
2015-11-19 & Ou8a &     49      &   300             &   0.86    &  0.922        & 9 & - & 44    & 5  \\
2015-11-27 & Ouka & 40  &   300             &   1.07    &  0.919        & 6 & - & 38  & 0  \\
2015-12-01 & Ouka & 39  &   300             &   1.17    &  0.920        & 8 & - & 41  & (3) \\
2015-12-14 & Ouka & 23  &   300             &   1.50    & 0.921 & 5 & - & 34  & (2) \\
\textbf{2015-12-18} & Ouka & 22 & 300 &   1.61  &  0.920        & 11   & \textbf{126} & \textbf{37}  & (3) \\

\end{tabular}
\end{table*}
}

\subsection{Data reduction}

For the 2015 recordings at the Ouka\"{i}meden Observatory, the spectroscopic data were reduced using the free and open-source astronomy software AudeLA dedicated pipeline \citep{KDM2012}, while the photometric data were processed using IRAF. Both packages perform classical operations, such as bias subtraction, flat-fielding, 
masking of bad pixels, wavelength calibration, and spectra extraction. The duration of our observations ranged from 1\,h 48\,min to 4\,h 07\,min, 
covering 13\% to 30\% of the 13.6\,h pulsation period, focusing on phases of rising light between 0.7 and 1.06 (see Table \ref{longobslog}). The wavelengths are given in the rest frame of RR\,Lyr considering a star velocity, $V_{*}$, with respect to the Solar System barycenter of -73.5\kms \citep{CG1996}.
Spectra were filtered with a Gaussian filter with a standard deviation of the distribution close to 2 pixels that keeps unchanged resolution power as resolution element is about 5.5 pixels at 5897\,\AA.
Knowing that the resolving power of the spectrograph is around 10,700 with a spectral dispersion of 0.1~\AA/px, this leads to a resolution of about five pixels for the D3 line, so this filter keeps the spectral resolution of our measurements. 

\subsection{Ephemeris computation}

In this paper, the pulsation phase is denoted $\varphi$, where $\varphi\equiv0.0$ corresponds to the luminosity maximum. 
The Blazhko phase is denoted $\psi$, where $\psi\equiv0.0$ when the highest luminosity amplitude is observed during a Blazhko cycle.
The period and the reference heliocentric Julian date (HJD) from a previous maximum both for pulsation and Blazhko cycles evolve with time. Thus updated parameters are required.

\cite{LPK2014} showed that the pulsation period varies alternately between two states, defined as a pulsation over long (longer than 0.56684\,d) and  short (shorter than 0.56682\,d) periods. These states alternate with intervals of 13--16\,y. After a sudden change in period value, this interval remains constant for a few years. The physical origin of these two states is not known yet.
The pulsation period, $P_P$, for spectra recorded in 1996 and 1997 were taken from Table\,3 in \citet{LPK2014}.  For 2013  and 2015, the $P_P$ values were computed using the difference between the two closest O--C minima from the GEOS RR Lyr database\footnote{The GEOS database is freely accessible online at \href{http://rr-lyr.irap.omp.eu/}{http://rr-lyr.irap.omp.eu/}.} and data close to our observations. The reference HJD maxima light HJD0$_\mathrm{P}$(max. light) for the 2015 were determined by photometry, carried out simultaneously with spectroscopy; whereas for other years, they were found in the GEOS database.

The variations of the Blazhko period $P_B$ of RR\,Lyr have been analyzed over more than one century by \citet{LPK2014}. At each interval of 1000-3000\,d, the value of $P_B$ jumps randomly by about a third of a percent. Thus, for a century, $P_B$ has varied between 38.8 and 41.2\,d, with the maximum value around 1975. 
These authors have shown that the pulsation and Blazhko periods changed in a way that is completely uncorrelated with each other.

Blazhko ephemeris were built using period $P_B$ from \cite{LPK2014}, while the reference HJD maximum light amplitude, hereafter HJD0$_\mathrm{B}$, were computed by finding the most important minima O-C values from the GEOS database by minization method.

Typical equation used for pulsation and Blazhko phase $E$ computation is:

\begin{equation}
  \mathrm{HJD(max.\,light)}=\mathrm{HJD0}+\mathrm{P} \times \mathrm{E.}
\end{equation}

 A summary of the ephemerides used for the different data sets is provided in Table~\ref{ephemeris}.

{\setlength{\tabcolsep}{4.4pt} 
\begin{table}[!ht]
\caption{Ephemerides used for the pulsation and Blazhko phase computation.}
\centering
\label{ephemeris}
\begin{tabular}{c|ll|ll}
\hline
\hline
Epoch & \multicolumn{2}{c}{Pulsation} & \multicolumn{2}{|c}{Blazhko} \\
\hline
      & HJD0$_{P}$              & Period                      & HJD0$_{B}$   & Period \\
      & {\scriptsize (-2\,400\,000)}& $P_P$ (d) & {\scriptsize (-2\,400\,000)} & $P_B$ (d)  \\ 
\hline
1994  & 49\,568.5063\tablefootmark{a} & 0.5668427\tablefootmark{b} & 49\,551.172\tablefootmark{a} & 39.03\tablefootmark{b} \\
1995  & 49\,937.5209\tablefootmark{a} & 0.5668427\tablefootmark{b} & 49\,903.701\tablefootmark{a} & 39.06\tablefootmark{b} \\
1996  & 50\,259.4761\tablefootmark{a} & 0.5668174\tablefootmark{b} & 50\,332.656\tablefootmark{a} & 39.06\tablefootmark{b} \\
1997  & 50\,599.5665\tablefootmark{a} & 0.5668174\tablefootmark{b} & 50\,679.550\tablefootmark{a} & 39.06\tablefootmark{b} \\
  2013  & 56\,578.450\tablefootmark{a} & 0.56683561\tablefootmark{b} & 56\,572.5\tablefootmark{a} & 39.0\tablefootmark{b} \\
  2015  & 57\,354,322\tablefootmark{a} & 0.56683561\tablefootmark{b} & 57\,195.734\tablefootmark{a} & 39.0\tablefootmark{b} \\ 
\hline
\end{tabular}

\tablefoot{Maximum light HJD and periods are from:
\tablefoottext{a}{GEOS;\,}
\tablefoottext{b}{\citet{LPK2014};\,}
}
\end{table}
}

\begin{figure}[!ht]
  \centering
  \includegraphics[width=1.00\hsize]{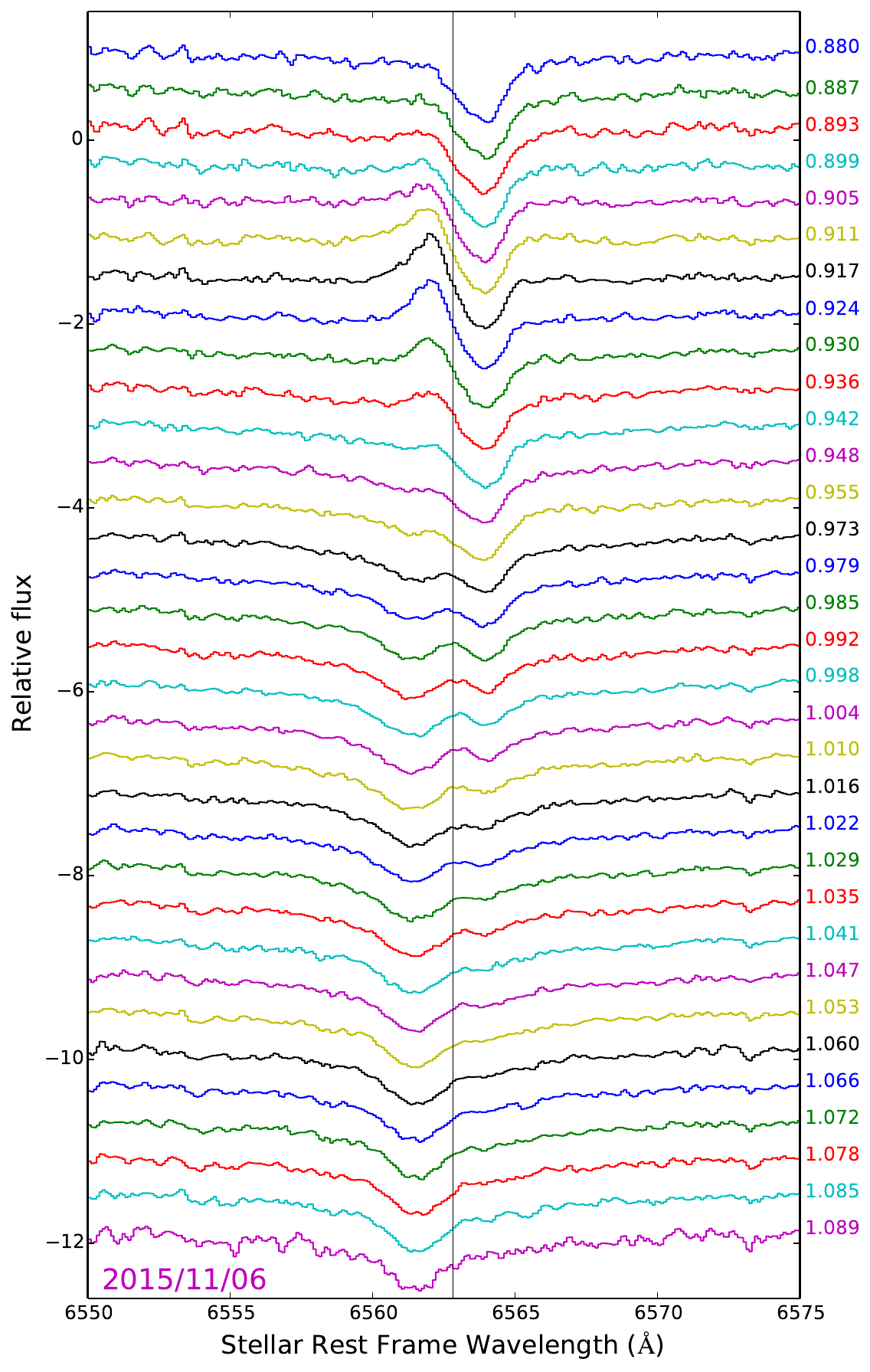}
  \caption{
    Evolution of the \ha line profiles of \rrlyr corresponding to night November 6, 2015. The pulsation phases are indicated on the right side and vary from $\varphi=0.880$ (top right) to $\varphi=1.089$ (bottom right), while the Blazhko phase $\psi$ at the beginning of the observing interval is 0.53. 
    The exposure time is 300\,s and weak absorption features are telluric lines.
    The vertical line represents the \ha line laboratory wavelength (zero velocity) in the stellar rest frame. The profiles are arbitrarily shifted in terms of flux.}
    \label{figure1}
\end{figure}

We can reasonably estimate that the uncertainty on $\rm HJD (max.\,light\,ampl.)$ is $\pm 2$ days and on the Blazhko period, $P_{B}$ is $\pm 0.2$ days. 

Based on the analysis of \citet{LPK2014}, the period of Blazhko cycle is 39.03\,d for HJD between 2\,448\,012 and 2\,450\,000 and 39.06\,d for HJD between 2\,450\,224 and 2\,452\,926.
\citet{Dal1995} assumes that the maximum julian day (October\,5, 1994) of the last ephemeris corresponds to the beginning of a new four-year Blazhko cycle. Consequently, all nights of observation obtained between 1995 and 1997 refer to the same four-year Blazhko cycle, while the nights of observations in 1994 belong to a preceding four-year Blazhko cycle.

Using the Blazhko ephemeris above, we calculated for each observed night, the HJD of the maximum light amplitude of light curve which is expected to occur just before the maximum light ($\varphi\equiv0.0$) of the observed pulsation cycle. This Blazhko HJD$_{\textrm{Max}}$ is given in Table \ref{longobslog}. Of course, this HJD is usually several Blazhko cycles after the one that was actually observed (ephemeris). This number of cycles is given by the integer part of the Blazhko phase $\psi$ of the Table \ref{longobslog}. The larger this number, the more uncertainty about $\psi$ increases. Deviations of up to 15\% or more can then occur.
For example with an extrapolation over 27 Blazhko cycles, the $\psi$ phases calculated for the nights of August 30 and 31, 1997 ($\psi = 3.30$ and 3.33, respectively, see Table \ref{longobslog}) could be affected by an error of 14\%.
On the other hand, to calculate $\psi$ for each night considered, we chose the date of the maximum luminosity of the pulsation cycle of the night even if it occurred early night or just after its end. In any case, since ten hours of observation (typical length of a night) only represents  a phase shift of $\Delta\psi=0.01$, the choice of the origin date for this calculation does not significantly affect the value of $\psi$ given in Table \ref{longobslog}.

\section{Observations of hydrogen in \rrlyn}

We analyzed the spectra recorded overnight in July and August 1994; August, September, and October 1995; June, July, and August 1996; May,  June,  and August 1997 -- for a total of 40 nights of observation totaling 1268 spectra. We found the presence of maximum photometric brightness on 19 nights out of the 40 nights. On these 19 nights, we observed in good conditions the presence of an emission within the H$\alpha$ line, but the study of his profile was possible only for 16 nights.

In 2015, we established an intensive observing campaign of the variable star RR\,Lyr that spanned over 51 days from October 29  to December 18. We found the presence 
of maximum photometric brightness on 10 nights out of a total of 40 nights of observation totaling 1129 spectra. During these 10 nights, we observed, in fair conditions, the main emission line of H$\alpha$ but the spectral profile was exploitable for only two nights. 

In Table \ref{longobslog}, we present all these characteristic of the \rrlyr observations made between 1995 and 1997 as well as in 2013 and 2015. In total, with the 40 spectra acquired on the night of October 12, 2013, we were able to examine 2437 spectra spread over 81 nights of observations spread over 21 years for the purposes of this study.

We analyzed the profile variations of \ha during all pulsation and Blazhko cycles of this set of observations. The core of the \ha line is formed far above the photospheric layers (hereafter, the upper atmosphere). 

\begin{figure}
  \centering
  \includegraphics[width=1.0\hsize]{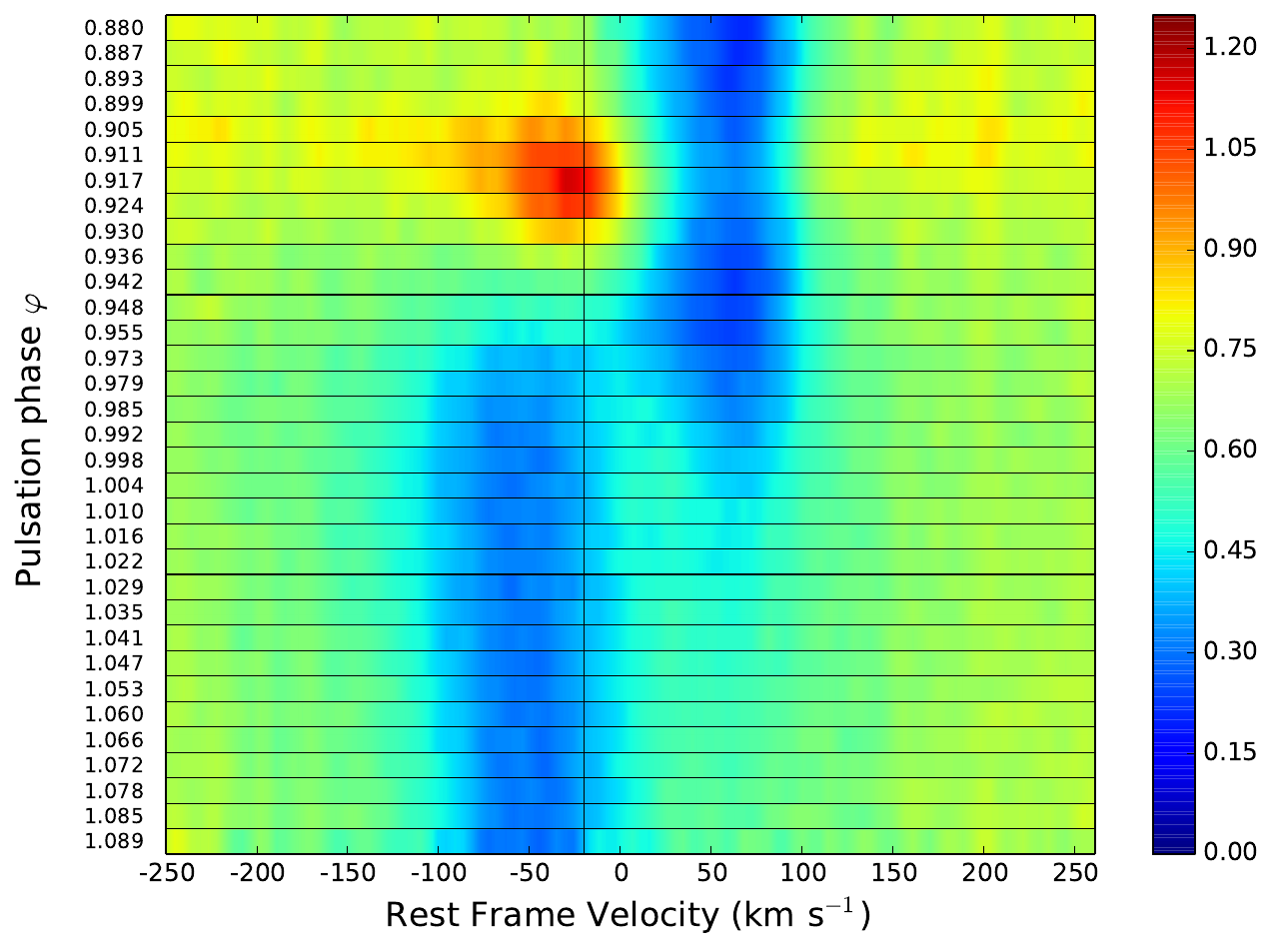}
  \caption{
    Time series of \ha line profiles of \rrlyr corresponding to night November 6, 2015 ($\psi=0.53$), are ordered by increasing phase from top to bottom. Spectra are interpolated to provide a 2-dimentionnal map of the \ha profile in the [-250, 250] \kms velocity range. Velocities are given in the stellar rest frame with positive velocities corresponding to inward motion (toward the photosphere). The pulsation phases $\varphi$ are given on the left side. On the right side, the color bar representing the flux is from 0.0 to 1.2.}
  \label{figure2}
\end{figure}

We present in Fig.~\ref{figure1} the 1D evolution of the \ha line profile between the pulsation phases $\varphi=0.880$ and 1.089 registered during the night of November 6, 2015, closed to a Blazhko minimum ($\psi=0.53$), namely, \,when the intensity of the main shock is in its least intense phase. The maximum of intensity is visible at $\varphi=0.917,$  followed by the line doubling phenomenon that was first interpreted by \citet{Sch1952}. In Fig.~\ref{figure2}, showing the 2D evolution of the \ha line profile, the opposite movement of atmospheric layers induced by this phenomenon is clearly visible through the inversion of the depths of the two \ha absorption components. Apart from the intensity of the emission lines and absorption lines, the 1D and 2D evolution of the \ha line profile is similar from minimum to maximum of the Blazhko cycle. Indeed, as shown by the comparison of Figs. 1 and 2 of \citet{GFL2013} and \citet{GSB2016} with our Figs.~\ref{figure1}  and \ref{figure2} of this article, the overall evolution of \ha remains the same.

\begin{figure}
  \centering
\includegraphics[width=1.0\hsize]{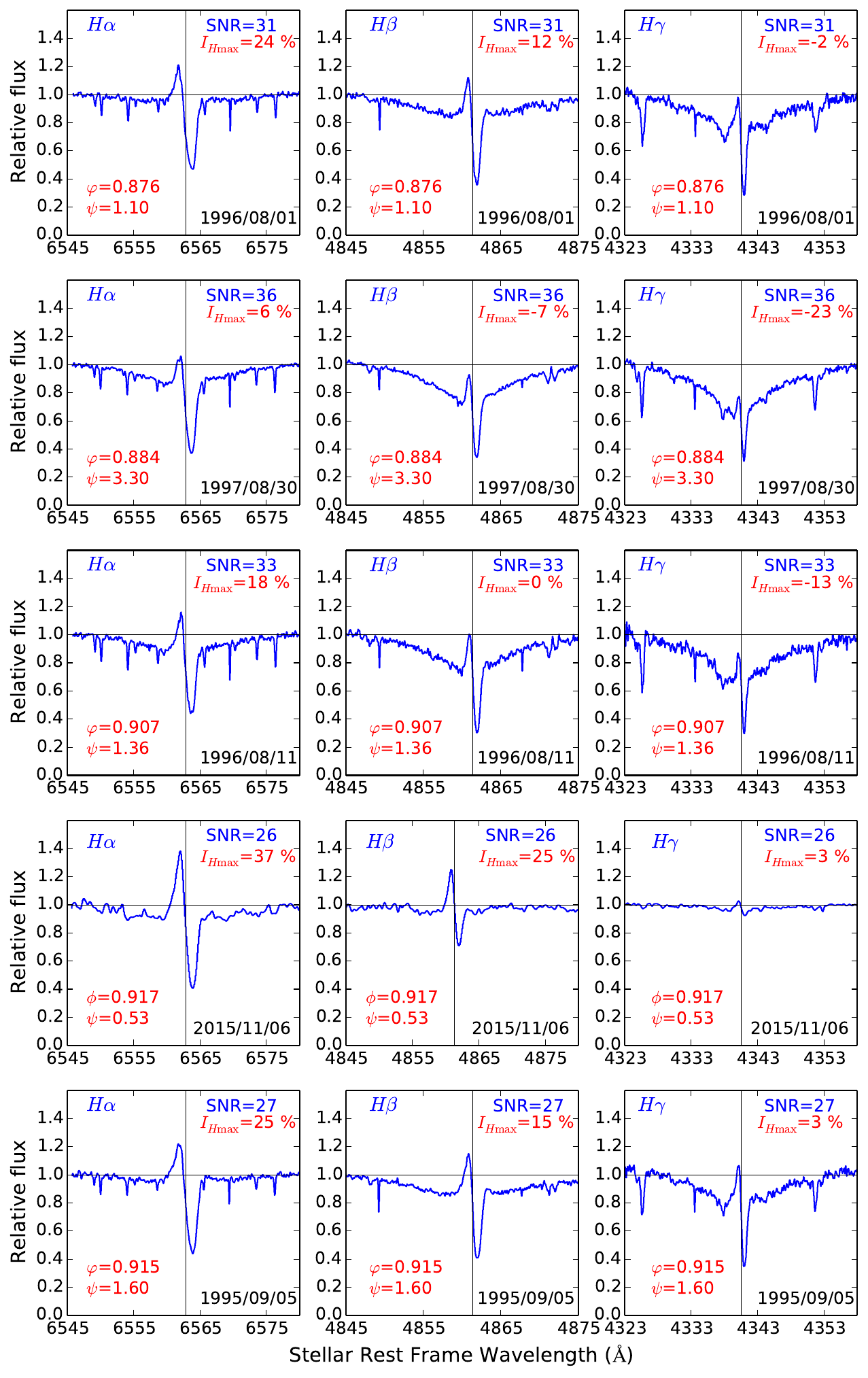}
  \caption{
    Line profiles of the three first members of the Balmer series for Blazhko phases just after the maximum ($\psi=0.10$) until just after the minimum ($\psi=0.60$) obtained around the pulsation phase 0.90 when these hydrogen emissions simultaneously reach their maximum intensity. The laboratory wavelength is marked by a vertical line and the continuum by a horizontal line.}
  \label{figure3}
\end{figure}

\begin{figure}
  \centering
\includegraphics[width=1.0\hsize]{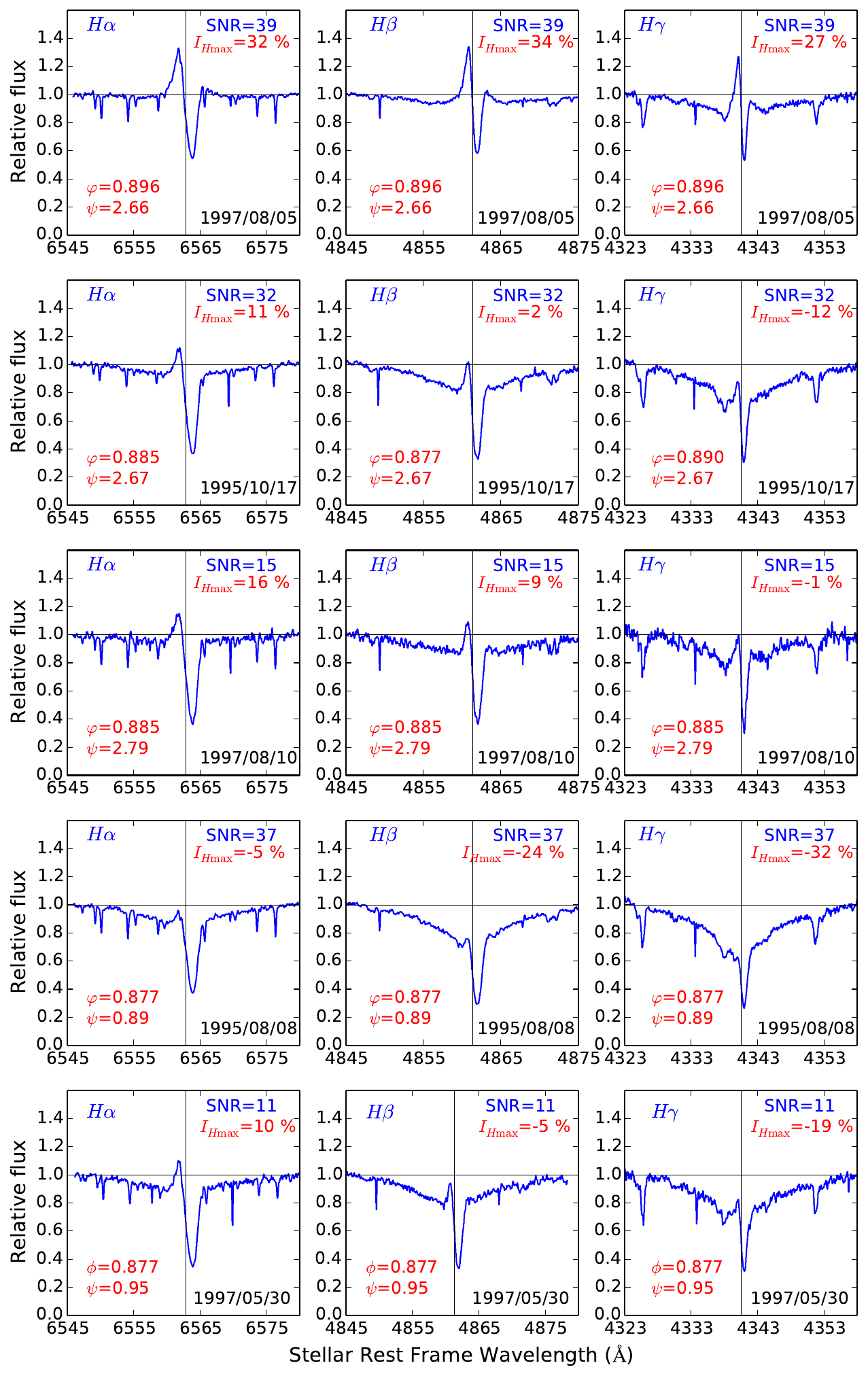}
  \caption{
    Same that Figure~\ref{figure3} but for Blazhko phases just after the minimum ($\psi=0.66$) until just before the maximum ($\psi=0.95$).}
  \label{figure4}
\end{figure}

Figures~\ref{figure3} and \ref{figure4} show the spectral profile of the \han, \hb, and \hg lines around the pulsation phase 0.90 when the hydrogen emissions simultaneously reach their maximum intensity. These lines have a complex profile. They are the result of a broad absorption line component produced within photospheric layers and a blueshifted emission line component produced within the shock wake that propagates at high speed within the atmosphere. The wide photospheric absorption is almost always affected by a Stark broadening which is mainly visible in the wings of the line. The hydrogen Stark broadening is minimum around the minimum Blazhko and maximum around the maximum Blazhko. It has been clearly established by several photometric studies that the average effective temperature is at its lowest level around the minimum Blazhko and at maximum around the maximum Blazhko. For example, this is the case on UZ Vir \citep{SHJ2012}, CZ Lac \citep{SJS2011}, or even MW Lyr \citep{JSS2009}; although we note that for this last star, the curve of the mean variation of $T_{eff}$ (their Fig.\,14) disagrees with their curves of variations of $T_{eff}$ versus the pulsation phase (their Fig.\,3). 
During the night of November 6, 2015 ($\psi=0.53$), the Stark broadening is practically absent at the minimum Blazhko (Fig.~\ref{figure3}). As the intensity of the emission decreases strongly from \ha to \hgn, the visibility of the Stark wings is more evident on this last line. 

As expected, the intensity gradually decreases from \ha to \hgn. However, due to the presence of large photospheric absorption, it is difficult to properly estimate the ratio of the intensities between these hydrogen emissions. Consequently, the magnitude of the deviations from LTE cannot be quantified since they must be significant as the emissions are produced in the radiative shock wake.

During the maximum of the expansion of the atmosphere, the intensity of the shock wave is subject to the Blazhko phenomenon because the intensity of the emission of the \ha line depends on the Blazhko phase \citep{PSP1965}. In 2015, the evolution of the \ha emission  registered during our ten pulsation cycles of observation presented a maximum of photometric emission at the Blazhko maximum. Nevertheless, while \ha was observed in emission at each pulsation cycle in 2015, the correlation between the Blazhko phase and the intensity of the \ha line is not always obvious when the S/N is weak.
Moreover, it appears that the intensity and the reproducibility of the Blazhko cycles are far from being constant. For example, in Figs.~\ref{figure3} and \ref{figure4},  although the precision on the determination of the Blazhko phases is limited (0.05), the intensity of the emissions of the hydrogen lines between the spectra at $\psi=2.66$ and $\psi=2.67$ varies between 20 and 40\%.

Concerning the study of the second and third emissions of \han, we limited our investigation to the \textsc{Elodie} spectra carried out at the OHP. Indeed, these two small emissions have a low average intensity of around 10\% with respect to the continuum level. First, the \textsc{Elodie} spectra have an average S/N that is three times greater than those performed at Ouka\"{i}meden Observatory. Second, the resolving power with \textsc{Elodie} is about 4 times better than that of Ouka\"{i}meden Observatory. Thus, with almost four times as many pixels to define profiles of such low intensity, the observations of \textsc{Elodie} offer a decisive advantage for characterizing these two small components in emission.

On the \textsc{Elodie} spectra recorded between the years 1994 and 1997, we observed the second and third emissions of the \ha line at all Blazhko phases. 
The limiting pulsation phases of the observation are between $\varphi=0.62$ and $\varphi=0.80$ for the second emission and between $\varphi=1.15$ and $\varphi=1.45$ for the third one. Depending on the night, it was possible to observe these emissions between 10 to 20\% of the pulsation cycle.

\section{Pulsation from the three \ha emissions}
\subsection{Shock velocity measurement}

As previously discussed by \citet{Gillet2019}, it is possible to obtain an estimate of the shock front velocity from the radial velocity of the blueshifted \ha emission during first apparition. The gas layers emitting the Balmer line radiations are located at the rear of the shock front in the hydrogen recombination zone (radiative shock wake). The shock models of \citet{FG2004} showed that the velocity inferred from Doppler shifts of Balmer lines is roughly one-third of the shock velocity:

\begin{equation}
V_{\textrm{shock}} \cong 3c\frac{(\lambda_{e}-\lambda_{0})}{\lambda_{0}}
\label{vshock1},
\end{equation}

\noindent 
where $\lambda_{e}$ is the wavelength of the maximum intensity (near $\varphi\simeq 0.90$) of the H$\alpha$ emission and $\lambda_{0}$ is its laboratory wavelength. This relation is mainly the consequence of the small optical thickness of the shock wake from the emitting Balmer layer, inducing a rapid postshock cooling. The Balmer emitted gas located just behind the front then undergoes strong compression and, therefore, slows down appreciably. When the shock velocity (represented by its Mach number, $M$) increases, the compression ratio quickly approaches of the limiting isothermal compression ratio $\gamma M^2$ where $\gamma = 5/3$ is the adiabatic exponent of a monoatomic gas. 

For the considered shock wave models, the validity of Eq.\,\ref{vshock1} is verified for shock velocities from 40\kms to 90\kmsn, namely, the velocity domain explored by \citet{FG2004}. All their models were computed for unperturbed atmospheric temperature $T_1 = 3000 $K. The values of $V_{\textrm{shock}}$ provided by the equation generate a dispersion of $\pm0.16V_{\textrm{shock}}$ compared to the exact values given by the models.

We applied this equation here to shock velocities up to 149\kms (Table \ref{longobslog}). Consequently, this extrapolation is questionable for velocities above 90\kms and our estimate of the shock velocity must be considered with caution. However, it is expected that the optical thickness of the shock wave in line frequencies still decreases when the shock velocity increases because of the increased ionization rate. 
Consequently, it is likely that Eq.\,\ref{vshock1} remains a satisfactory approximation in the shock velocity domain considered here.

\begin{figure}[!ht]
  \centering
  \includegraphics[width=1.00\hsize]{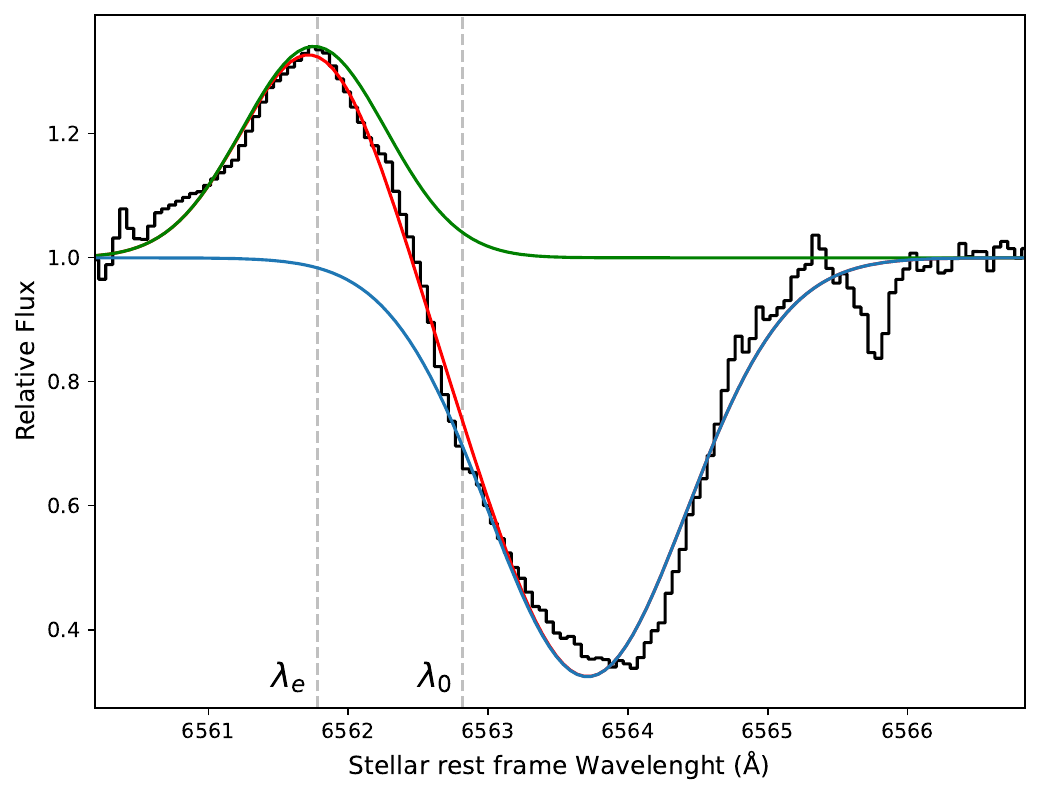}
  \caption{Example of the most plausible fitting for 1996-08-01 spectrum (in black). The synthetic spectra of emission and absorption components are respectively ploted in green and blue color whereas the sum is ploted in red. Note: the distorted red wing of the emission component. The wavelength of the emission component, $\lambda_e$, is measured on the synthetic spectrum of the emission component (green color), thus managing the blueshifting described in the text.}
  \label{figure5}
\end{figure}

In order to estimate the emission component's wavelength, $\lambda_{e}$, we made use of the inverse problem approach. We computed synthetic spectral profiles $F$ by adding a constant and two Gaussian functions of the wavelength. These quantities are defined by a seven-component parameter,
$p$: the constant continuum level for the whole profile and, for each Gaussian function, the intensity, the sigma, and the line center.
The inverse problem consists in minimizing the least-squares misfit function $J(p)=\int_{\Lambda}(F(p)-O)^{2}d\lambda,$ where $O$ is the recorded spectrum.
The interval $\Lambda$ is a limited wavelength zone large enough to include the studied line and but not too wide, thereby saving on long computation times. We note that the sum of these two gaussians allows to take into account the distortion of the red wing of the emission by the absorption one, as shown in Fig.~\ref{figure5}. 

We are not only interested in the optimal parameter, $\tilde{p}$, which gives the optimal fit, but also to the uncertainties associated with such a determination. The most plausible parameters are those
that lead to a small increase of $J(\tilde{p})$.
Based on the S/N values, we founded that such an increase must not exceed
5\%. Thus, contrary to Monte Carlo methods, we systematically explored
the whole set of parameters that match this criterion and selected the
extremal values for the emission component's wavelengths,
$\lambda_{e}^{min}$ and $\lambda_{e}^{max}$. From there, we deduced
the uncertainty on the shock wave velocity, hereafter $0.5(V_{s}^{max}-V_{s}^{min}),$ and
the asymmetric vertical bars' length (as used in Fig.~\ref{figure5}) that are then $+(V_{s}^{max}-V_{s})$ and $-(V_{s}^{min}-V_{s}); $
here, $V_{s}$ is the shock wave velocity associated with the center
of the emission component of the optimal parameter, $\tilde{p}$. More
details can be found in a forthcoming paper (Lailly et al., in prep.).

\medskip

Since the \ha emission is blueshifted when the main shock wave crosses the atmosphere, it is placed in the blue wing of the large photospherical absorption line. This positioning has the consequence to slightly ''blueshift'' by a few \kms the maximum of the emission.
Our method makes it possible to take this blueshifting into account because $\lambda_e$ is measured on the single synthetic spectrum of the emission component. Moreover, we have estimated that the difference between $V_s$ estimated by the double fit and by a simple Gaussian was around 2.3 \kmsn.

\subsection{Shock velocity versus Blazhko phase}

\begin{figure}[!ht]
  \centering
  \includegraphics[width=1.0\hsize]{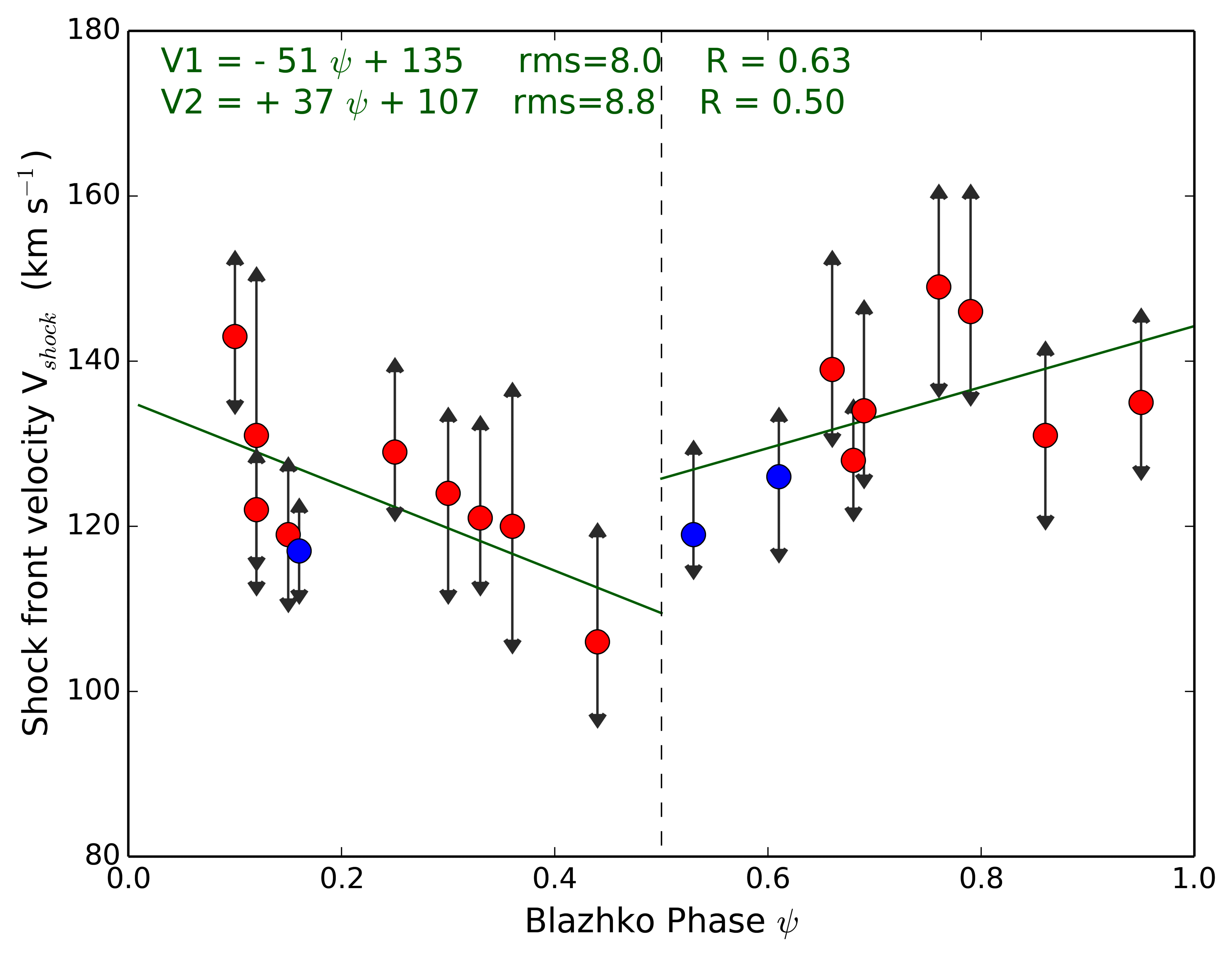}
  \caption{Shock wave velocity estimated from Eq.\,\ref{vshock1} at different Blazhko phases. The red dots refer to the 90s and the blue dots refer to the years 2013 and 2015. The uncertainty on the shock velocity induced by using Eq.\,\ref{vshock1} is certainly not smaller than 10 \kms when the shock velocity varies respectively from 100 to 150 \kmsn. The uncertainty on the Blazhko phase is estimated at 0.1. The asymmetric vertical bar through each point represents the range of plausible values of $V_s$ used to find the best fit. The straight linear regression lines through the points between $0.0\leqslant\psi\leqslant0.5$ and $0.5\leqslant\psi\leqslant1.0$ are shown. Their correlation coefficients $R$ are 0.63 and 0.50, respectively.}
  \label{figure6}
\end{figure}

The values of the maximum velocity of the main shock wave occurring around the pulsation phase 0.90 given in Table \ref{longobslog} are presented in Fig.~\ref{figure6} according to the Blazhko phase. The straight linear regression lines through the points between $0.0\leqslant\psi\leqslant0.5$ and $0.5\leqslant\psi\leqslant1.0$ are shown Fig.~\ref{figure6}. Their correlation coefficients $R$ are 0.63 and 0.50 respectively. It is clear that the large dispersion of the points strongly degrades the correlation. 

Although the dispersion of the velocities is appreciable, it seems that the shock velocity is maximum at the maximum Blazhko and minimum at the minimum Blazhko. The dispersion is linked both to inaccuracies within observations used but probably also to irregularities occurring from one pulsation cycle to the next due to the intensity of different physical processes (shock deceleration, gravitational infalling motion, shock radiative losses, etc.) occurring in the atmosphere. 
This type of irregularity linked to the dynamic response of the atmosphere when the shock crosses the latter, was  previously noticed by \citet{Gil2013} in the evolution of the light curve amplitude versus the pulsation cycle number during: the first quarter of Kepler observations (Q1) conducted between 2009 May 16 and June 10 and during the second quarter of Kepler observations (Q2) conducted between 2009 July 4 and August 11 (see his Figs.\,2 and 3).

Finally, despite the fluctuations in the shock velocity from one pulsation cycle to another and the irregularities of the Blazhko cycle, it can therefore be established that a shock velocity decreases on average from a Blazhko maximum to a next minimum and then increases again from a minimum to a maximum.
It is also likely that a jump of about 20\,\kms will occur when the shock velocity starts to increase again (Fig.~\ref{figure6}).
Thus, on average, a shock velocity increases from 100 to 150\kms, which would induce an appreciable increase in the kinetic energy of the shock and therefore of the maximum radiative flux produced by the latter, which is proportional to the third power of the shock velocity \citep{Gillet2019}.

It would therefore appear (as an observational result) that the shock velocity tends to gradually increase between the Blazhko minimum and its maximum. We consider what physical mechanism could explain such a process. Furthermore, when the Blazhko maximum is reached, we consider what makes the shock velocity suddenly start to decrease and what makes it start to rise again from the Blazhko minimum.

\begin{figure}[!ht]
  \centering
  \includegraphics[width=1.0\hsize]{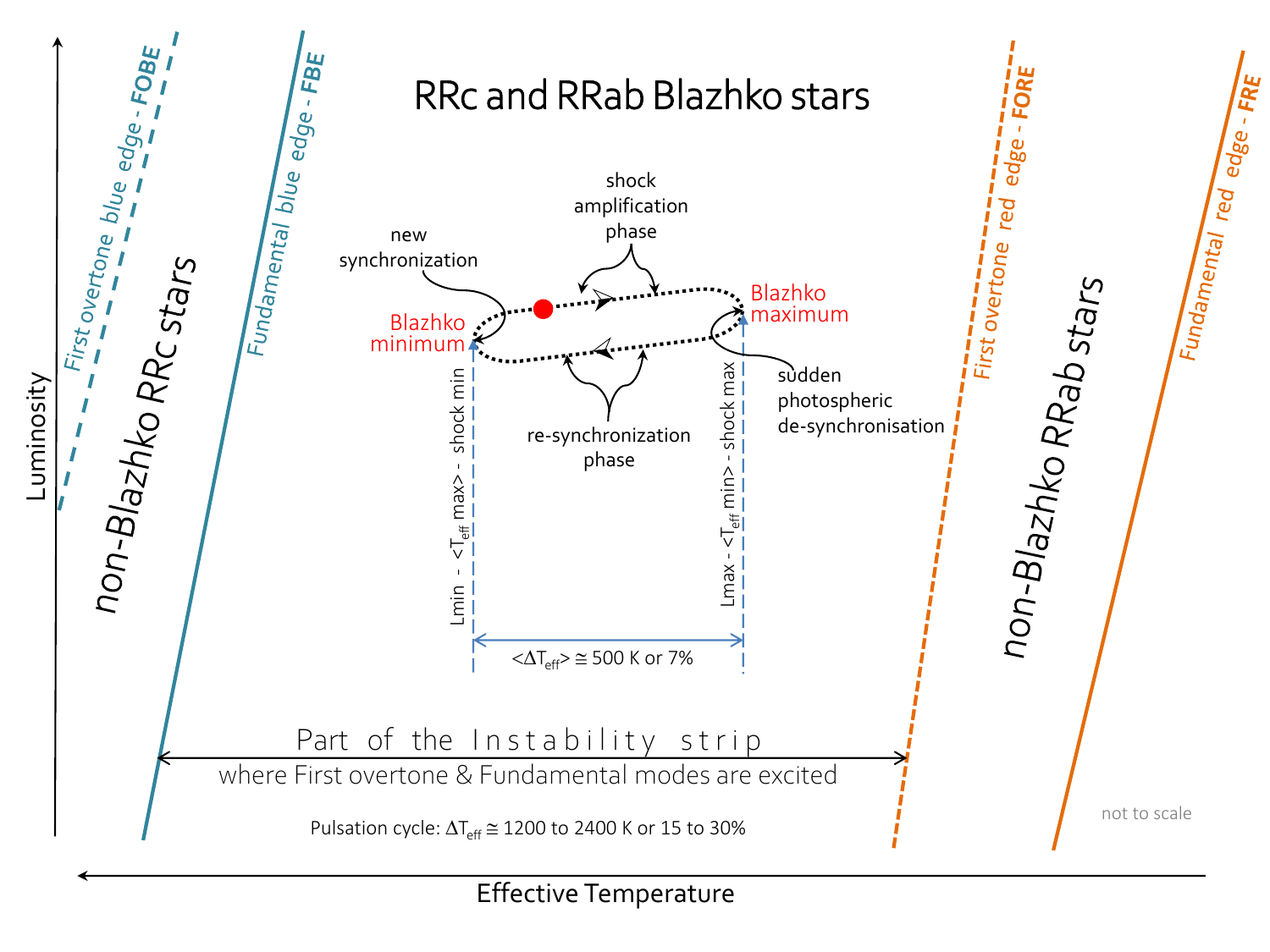}
  \caption{Different mechanisms of atmospheric dynamics occurring during the Blazhko cycle. They cause rapid loops of a few tens of days (Blazhko period) with 
an appreciable amplitude around 500\,K and a few solar luminosities when the star in Blazhko phase crosses the instability band of the Herztsprung-Russell diagram. 
The fundamental mode edges of the instability strip are plotted as solid lines and first overtone mode edges as dashed lines.}
  \label{figure7}
\end{figure}

Based on a series of nonlinear nonadiabatic models of RR\,Lyr that allow for the propagation of shock waves in its atmosphere calculated by \citet{FG1997}, \citet{Gil2013} proposed a new explanation of the Blazhko phenomenon. For RR\,Lyr, he notices that in addition to the fundamental mode, the first overtone is probably excited marginally during each pulsation cycle. This overtone induces at each pulsation cycle a shock of moderate intensity which has the effect of slowing down the fallout of the high atmospheric layers and, consequently, attenuates the amplitude of the $\kappa$-mechanism. Observations show that when the star approaches the Blazhko maximum, its average $T_{eff}$ decreases, thus moving the star away from the optimum excitation region of the first overtone (Fig.~\ref{figure7}). As a result, the atmospheric braking is reduced and the intensity of the $\kappa$-mechanism increases at the same time as the intensity of the main shock. Therefore, in cycle after cycle of pulsation, there is an (irregular) amplification of the intensity of the shock. After a while, the shock becomes powerful enough to desynchronize the pulsating movement of the atmospheric layers. This sudden break occurs (by definition) at the maximum Blazhko. Then, the photospheric layers relax dynamically before returning to their state of synchronous pulsation at the Blazhko minimum (by definition). During this phase of relaxation, the main shock being of lower intensity ($V_{\textrm{shock}}$ decreases), it loses less energy by radiation, and consequently, the average $T_{eff}$ increases again gradually to reach a new maximum at the Blazhko minimum.
When the synchronous state is reached, the intensity of the main shock can again increase (as the value of  $V_{\textrm{shock}}$) and the mean $T_{eff}$ decreases. Thus, a new Blazhko cycle starts again. Figure~\ref{figure7} summarizes these different mechanisms of atmospheric dynamics occurring during the Blazhko cycle. These mechanisms end up causing loops of appreciable amplitude around 500\,K and a few solar luminosities when the star in Blazhko phase crosses the  instability band of the Herztsprung-Russell diagram.

Currently, the systematic excitation of the first overtone at each pulsation cycle is not yet definitively established, as its possible effect on atmospheric dynamics. However, the signature of this mode in RR\,Lyr has already been found in the Q5–Q6 Kepler photometric data \citep{MKS2012}. Nevertheless, the first overtone has a very small amplitude (few millimagnitudes), in contrast with the fundamental mode, explaining its non-detection from the ground observations. The radial nonlinear hydrodynamic calculations of these authors demonstrated that the fundamental mode can lose its stability due to its interaction with the first radial harmonic. In addition,
although the number of Blazhko stars used is limited, \citet{Gil2013} showed that all are within an intermediate region of the Instability Strip in the Herztsprung-Russell diagram in which both the fundamental and the first overtone modes of pulsation are excited at the same time. In contrast, the location of non-Blazhko stars is redder and closer to the fundamental mode red-edge than that
of Blazhko stars (Fig. 1 of \citet{Gil2013}). Thus, non-Blazhko stars would not be affected at all (or only slightly) by the first overtone during their pulsation cycle. This general trend needs to be confirmed by more detailed observations. 
The first overtone signals were also identified in three other RR\,Lyrae stars in the Kepler sample by \citet{BKS2010}. More generally, Blazhko stars of RRab type 
(of large amplitudes) systematically show the presence of additional modes especially the first harmonic that would modulate the fundamental mode; whereas non-Blazhko 
RRab stars never do \citep{BS2015}. All these results show that further theoretical studies still need to be carried out in this direction. 

Finally, the explanation of the shock model of the Blazhko effect \citep{Gil2013} seems quite plausible and has the merit of highlighting the possible key role that hypersonic shock waves could play on atmospheric dynamics by significantly affecting the pulsation as is well observed within Blazhko stars.
 This is consistent with the considerable shock wave intensity and its important radiative flux around the Blazhko maximum ($\psi = 0.0$). Indeed, depending on the physical and dynamical conditions occurring in the atmosphere, this flux can be equal to and even greater than that produced by the photosphere.

\begin{figure}[!ht]
  \centering
  \includegraphics[width=1.0\hsize]{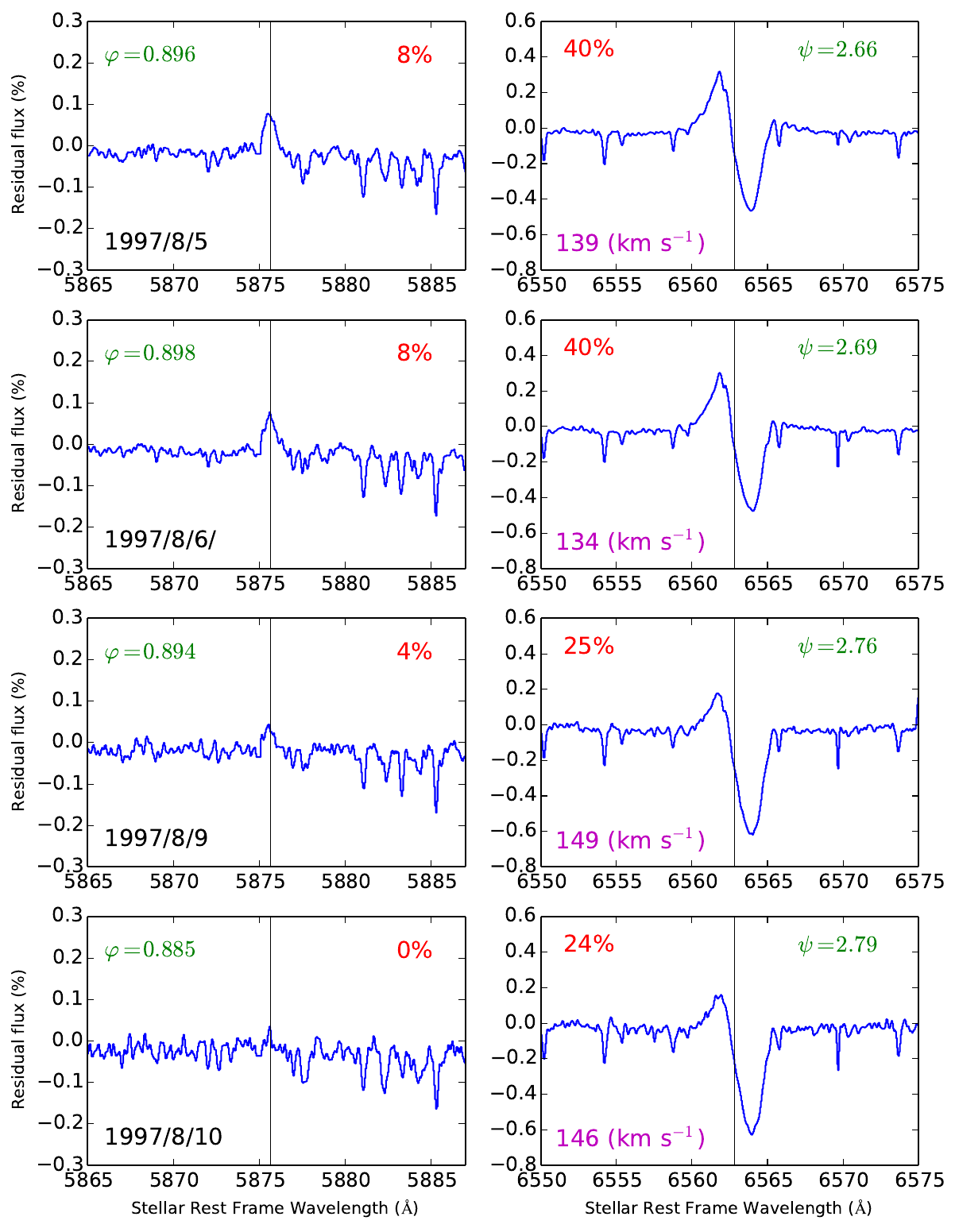}
  \caption{Line profiles of the maximum emission of D3 and H$\alpha$ at Blazhko phase corresponding to nights August 5, 6, 9, and 10, 1997. For each pair of observation (D3 and H$\alpha$), the Blazhko phase $\psi$, the pulsation phase $\varphi$ and the intensity of each emission is given with respect to the continuum. The shock velocity is shown in the lower left corner of the right column. The laboratory wavelength is marked by a vertical line. The very narrow "emission" occurring at the wavelength of D3 on August 10, 1997 is probably caused by noise.}
  \label{figure8}
\end{figure}

Our observations seem to show that the velocity of the shock wave gradually increases from the Blazhko minimum to its maximum then, conversely, decreases from the maximum to the next minimum. However, as seen in Fig.~\ref{figure6}, a jump in the shock velocity of the order of 20 \kms seems probable just after it has reached its minimum value at $\psi=0.5$. Although this fact needs to be confirmed by other finer observations, this jump would highlight the large energetic efficiency of the process of synchronization of the atmospheric layers.

Moreover, as highlighted by \citet{Gil2013}, the evolution of the energy (or the velocity) of the shock from one pulsation cycle to the next is not monotonic. It presents important fluctuations certainly caused by the nonlinear coupling of the ballistic movement of the highest atmospheric layers with the photospheric ones. It is clear that the use of Eq.\,\ref{vshock1} also induces an appreciable dispersion of the estimation of the shock velocity (see Fig.~\ref{figure6}) because it results (on the one hand) from an approximate observation of the results of the shock models and (on the other hand) the models remain limited because, for example, they only consider pure hydrogen gas \citep{FG2004}. In order to confirm this cyclical evolution of the shock velocity during the Blazhko period, it would be useful to be able to confirm this variation with new high quality observations that would require a telescope (with at least
3\,m in diameter ) to be available for several months. If confirmed, this would constitute a major observational clue depending on the explanation of the Blazhko effect proposed by \citet{Gil2013}.

\subsection{Decay of maximum intensity of emission lines}

Figure~\ref{figure8} shows the evolution of the maximum intensity of the first and main emission of the H$\alpha$ line, as function of the Blazhko phase, obtained during the nights of August 5, 6, 9, and 10, 1997, corresponding to the same Blazhko cycle ($\psi=2.66$, $\psi=2.69$, $\psi=2.76,$ and $\psi=2.79$ respectively). These observations were carried out on a time base of 5 days (13\% of the Blazhko period which was 39 days in 1997 \citep{LPK2014} and started just from a minimum Blazhko. They were spread over nine pulsation cycles ($P=0.56$\,days). It can be seen that the maximum emission intensity of the H$\alpha$ line is 40\% just after the Blazhko minimum phase $\psi=2.66-2.69$ and it decreases almost by a factor of 2 (24\%) at the Blazhko phases $\psi=2.76-2.79$. These observations establish that when the shock velocity increases, the maximum intensity of the emission lines of hydrogen and helium decreases from the Blazhko minimum. This unexpected result is quite novel.
It is certainly a consequence of the increase in the ionization rate of hydrogen and helium when the shock velocity increases. Indeed, the emission lines of hydrogen and helium are formed in the radiative wake of the shock. The latter is a very narrow region located behind the shock front whose very high temperature decreases rapidly. Thus, unlike the unperturbed atmosphere, these emission lines can only be produced within the radiative shock wake.

\begin{figure}[!ht]
  \centering
  \includegraphics[width=1.00\hsize]{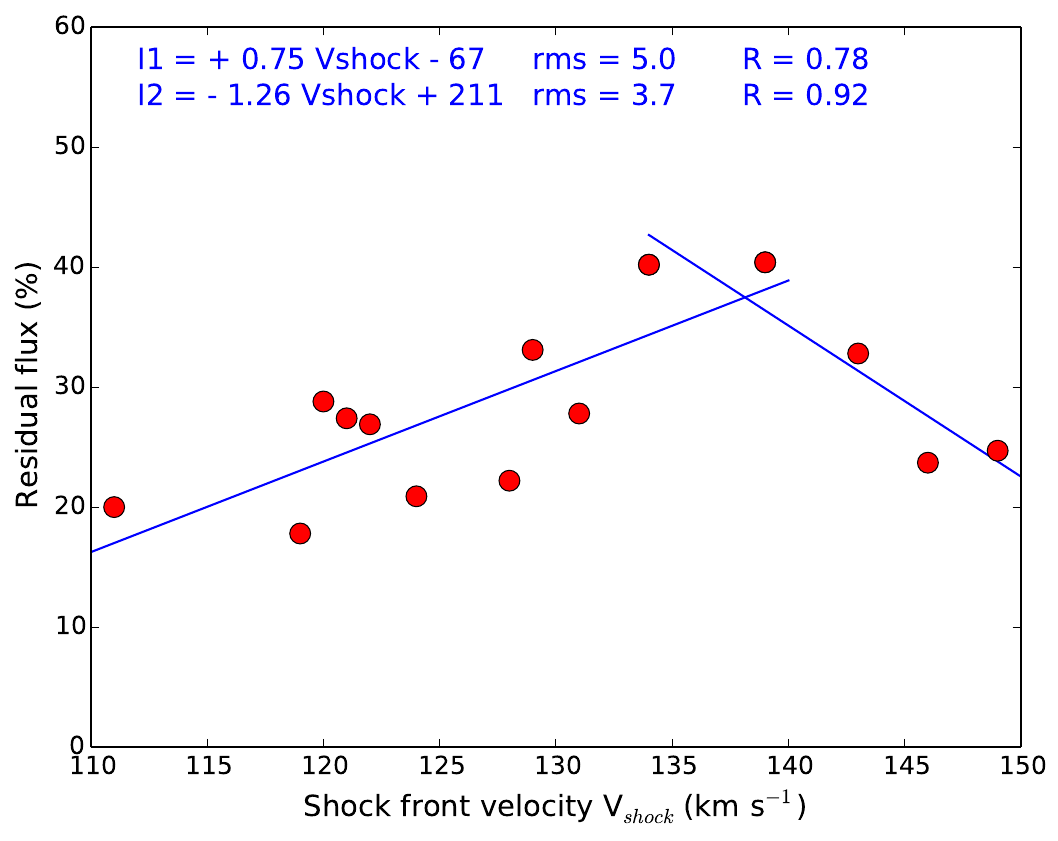}
  \caption{Maximum intensity of the first and main emission of the H$\alpha$  versus the corresponding shock velocity (red dots: 1990s, blue dots: 2015; see Table \ref{longobslog}). Mean uncertainty of residual 
  flux is 2\%. The straight linear regression lines through the points between $110\leqslant V_{shock} \leqslant 135$ and $135\leqslant V_{shock} \leqslant 150$ are shown. 
The uncertainty on the shock velocity induced by using Eq.\,\ref{vshock1} is certainly not smaller than 10 \kms when the shock velocity varies respectively from 100 
to 150 \kmsn.}
  \label{figure9}
\end{figure}

In the previous section, we show that the maximum shock velocity during a pulsation cycle increases from the Blazhko minimum to its maximum. As a result, the ionization rate in the wake also increases with the value of the shock velocity. There will therefore come a time when the number of neutral hydrogen atoms available in the wake will no longer allow the emission intensity of the line to be increased. Accordingly, there will be an appropriate value of the shock velocity, for which the intensity of \ha and D3 will reach a maximum. The measurements that we carried out on the observations analyzed in this work (see Table \ref{longobslog}) suggest that this peak of intensity occurs around $V_s\simeq 137$\kms (see Fig.~\ref{figure9}).

It should be noted that we opted for an absence of emission of D3 on August 10, 1997 despite the presence of a small intensity peak of about 3\% of the continuum level at the wavelength of D3 (see Fig.~\ref{figure6}). The full width at half maximum (FWHM) of D3 observed on August 5, 6, and 9, 1997 was around 43\kms and on August 10, 1997 it was reduced to 7\kmsn. These values correspond respectively to thermal widths of 40\,000\,K and 1000\,K. The latter value is far too low to correspond to the atmospheric temperature occurring in this type of star. Moreover, since it is identical to the width of the other emission fluctuations present in the spectrum, we conclude that this very narrow "emission" occurring at the wavelength of D3 is, in fact, caused by noise -- and not by helium.

A similar ionization phenomenon makes it also possible to explain the disappearance of the Fe\,I ($\lambda 4934.006$) line around $\varphi=0.96$, in favor of an increasing the intensity of the Fe\,II ($\lambda 4923.921$) absorption line during the passage of the shock wave. This Fe\,I absorption line reappears from $\varphi=1.00,$ that is to say, at the end of the passage of the shock wave in the RR\,Lyr photosphere. This phenomenon has already been observed by \citet{CVG2008} on another RR\,Lyrae type star, namely, S\,Area.

\begin{figure}[!ht]
  \centering
  \includegraphics[width=0.97\hsize]{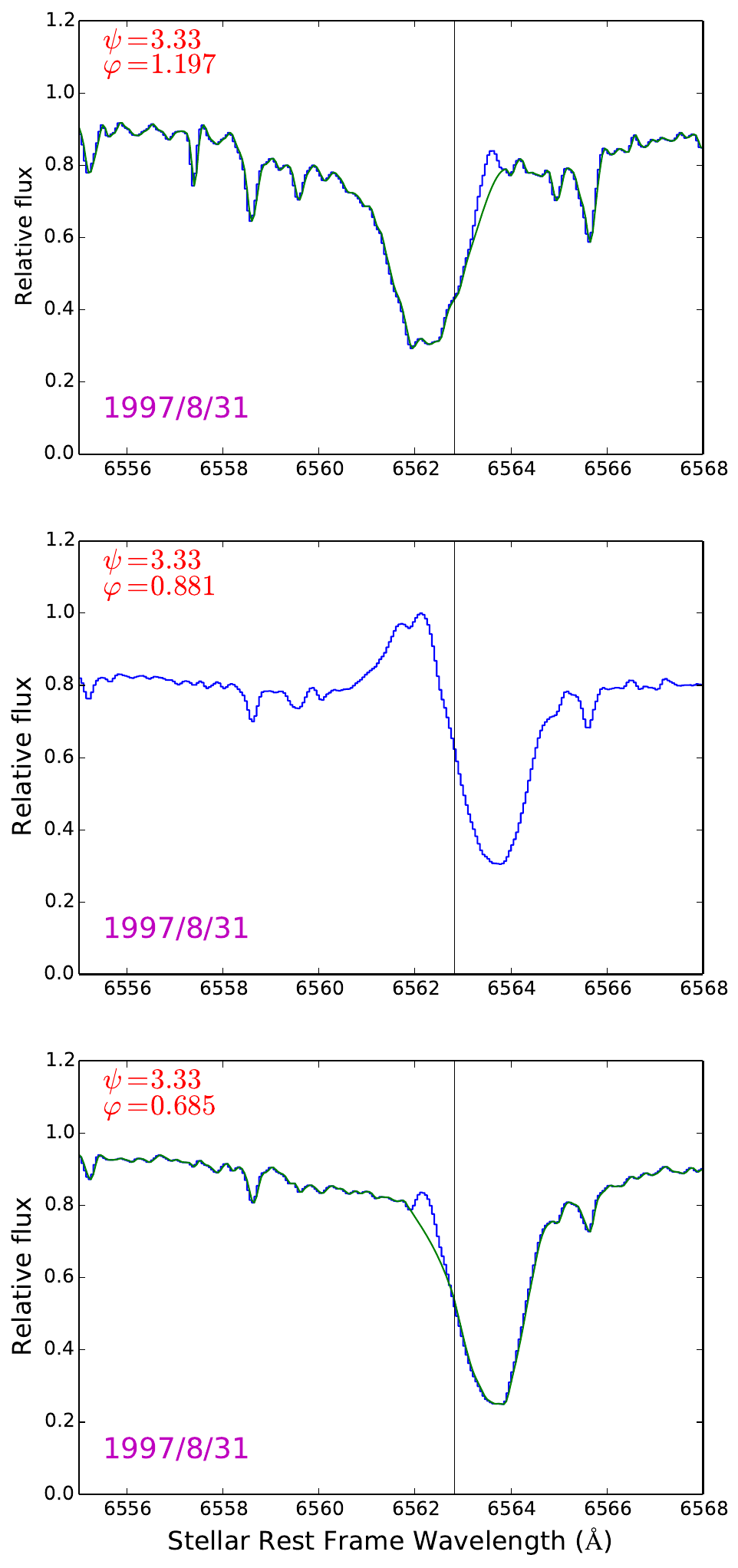}
  \caption{Second (up), first (middle), and third (down) apparitions of the emission that occurred in the large photospheric absorption profile of H$\alpha$ during the night of August 31, 1997. The fitting of the spectrum makes it possible to highlight the two smallest emissions. The Blazhko phase, $\psi,$ and the pulsation phase, $\varphi,$ are given in the left upper corner.The laboratory wavelength is marked by a vertical line.}
  \label{figure10}
\end{figure}

\subsection{Occurrence of the second and third appearances of the hydrogen emission line}

{\setlength{\tabcolsep}{4.5pt}
\begin{table*}
\centering
\caption{Occurrence of the \ha second emission of RR\,Lyr around the pulsation phase $\varphi_{2Max}$ observed with ELODIE between 1994 and 1997.
Columns provide the date of the night, the corresponding Blazhko phase $\psi$, the pulsation phases at the beginning and at the end of the emission and at its maximum intensity. For this latter, the corresponding blueshifted velocity (\kmsn) and the intensity of the H$\alpha$ second emission with respect to the continuum are $V_{2Max}$ and $I_{2Max}$, respectively. The last column gives the shock velocity, $V_{shock}$ (\kmsn).}
\label{Ha2}
\begin{tabular}{ccccccccc}
\hline \hline
Night & $\psi$ & S/N  & $\varphi_{2start}$ & $\varphi_{2end}$ & $\varphi_{2Max}$ & $V_{2Max}$ & $I_{2Max}$ & $V_{shock}$ \\
{\scriptsize (yyyy-mm-dd)} &   &   &   &  & & (\kms)       & (\%) & (\kms) \\
\hline 
1994-07-27 & 0.26       & 10 & 0.651    & 0.739 & 0.676 & 29  & 8 & - \\
1994-08-03 & 0.44       & 11 & 0.779    & 0.805 & 0.779 & 24  & 8 & 106 \\
1995-08-08 &  0.89      & 11 & 0.658    & 0.725 & 0.712 & 24  & 7 & 131 \\
1995-09-06 &  1.63      & 11 & 0.681    & 0.767 & 0.707 & 21  & 14 & - \\
1996-08-11 & 1.36       & 11 & 0.616    & 0.776 & 0.678 & 23  & 13 & 120 \\
1997-08-07 & 2.72       & 10 & 0.681    & 0.736 & 0.707 & 26  & 8 & - \\
1997-08-31 & 3.33       & 14 & 0.685    & 0.711 & (0.685) & 23  & 10 & 121 \\
\end{tabular}
\end{table*}
}

{\setlength{\tabcolsep}{4.5pt}
\begin{table*}
\centering
\caption{Occurrence of the \ha third emission of RR\,Lyr around the pulsation phase, $\varphi_{3Max}$, observed with ELODIE between 1994 and 1997.
Columns provide the date of the night, the corresponding Blazhko phase $\psi$, the pulsation phases at the beginning and at the end of the emission and at its maximum intensity. For this latter, the corresponding blueshifted velocity (\kmsn) and the intensity of the H$\alpha$ third emission with respect to the continuum are $V_{3Max}$ and $I_{3Max}$, respectively. The last column gives the shock velocity $V_{shock}$ (\kmsn).}
\label{Ha3}
\begin{tabular}{ccccccccc}
\hline \hline
Night & $\psi$ & S/N  & $\varphi_{3start}$ & $\varphi_{3end}$ & $\varphi_{3Max}$ & $V_{3Max}$ & I$_{3Max}$ & $V_{shock}$ \\
{\scriptsize (yyyy-mm-dd)} &   &   &   &  & & (\kms) & (\%) & (\kms) \\
\hline 
1994-07-28 & 0.29       & 8 & 1.183     & 1.372 & 1.216 & 43  & 11 & - \\
1994-07-29 & 0.31       & 7 & 1.194     & 1.252 & 1.240 & 41  & 11 & - \\
1994-08-05 & 0.49       & 10 & 1.200    & 1.453 & 1.375 & 40  & 9 & - \\
1995-09-05 &  1.60      & 7 & 1.219     & 1.375 & 1.282 & 44  & 14 & - \\
1996-06-24 & 0.12       & 10 & 1.157    & 1.221 & 1.210 & 34  & 11 & 122 \\
1996-06-26 & 0.18       & 9 & 1.306     & 1.373 & 1.317 & 33  & 11 & - \\
1996-07-31 & 1.07       & 10 & 1.154    & 1.353 & 1.252 & 40  & 10 & - \\
1996-08-01 & 1.10       & 9 & 1.157     & 1.306 & 1.306 & 41  & 10 & 143 \\
1996-08-09 & 1.30       & 9 & 1.173     & 1.348 & 1.267 & 37  & 9 & - \\
1996-08-25 & 1.71       & 10 & 1.238    & 1.408 & 1.335 & 44  & 11 & - \\
1996-08-27 & 1.76       & 8 & 1.195     & 1.213 & 1.213 & 37  & 7 & - \\
1997-08-05 & 2.66       & 7 & 1.192     & 1.326 & 1.299 & 40  & 8 & 139 \\
1997-08-07 & 2.72       & 10 & 1.304    & 1.391 & 1.347 & 39  & 9 & - \\
1997-08-08 & 2.74       & 9 & 1.196     & 1.395 & 1.299 & 37  & 7 & - \\
1997-08-09 & 2.76       & 7 & 1.263     & 1.363 & 1.334 & 37  & 9 & 149 \\
1997-08-29 & 3.28       & 12 & 1.181    & 1.350 & 1.243 & 44  & 9 & - \\
1997-08-30 & 3.30       & 10 & 1.217    & 1.317 & 1.272 & 46  & 9 & 124 \\
1997-08-31 & 3.33       & 5 & 1.153     & 1.197 & (1.197) & 42  & 8 & 121 \\
\end{tabular}
\end{table*}
}

Figure~\ref{figure10} shows the first, second and third emissions that occurred in the large photospheric absorption profile of H$\alpha$ during the same night of August 31, 1997, just after a Blazhko maximum ($\psi=3.33$). It was the only night in our sample to feature these three emissions consecutively. For this night, the observations starts at the pulsation phase 0.685 and ends at 1.197. The maximum of the main emission (first apparition) occurs at $\varphi = 0.881$ with a moderate intensity of 27\% which corresponds to a moderate shock velocity of 121\kmsn. This is consistent with the fact that the emission of He\,I $\lambda 5875.66$ line (D3) is not detected (Table \ref{longobslog}). 
The second emission appears on 4 successive spectra between $\varphi = 0.685$ and $\varphi = 0.711$. The emission with the maximum intensity is on the first spectrum of the night, which does not preclude the possibility that the actual maximum emission did not occur before $\varphi=0.685$. 

The third emission is also observed on four consecutive spectra between phases 1.161 and 1.197. The most intense emission occurs on the last spectrum of this series at phase 1.197. As this spectrum is the last realized of the night, the maximum observed intensity may not yet be at its effective maximum. In fact, we are observing the start of the visibility phase of the third emission, which should end around $\varphi=1.4$.

During this night, the second emission was observed during 2.6\% of the pulsation period and the third emission during 3.6\%. The first and main emission is observed between $\varphi = 0.846$ and $\varphi = 0.910$ (6\% of the period) with a maximum intensity at $\varphi = 0.881$.

In fact, although these three emissions occur during the same observing run, the third and second are the consequence of the main shock induced by the previous cycle which is at the origin of the ballistic movement in progress (Fig.~\ref{figure11}). These two small emissions are directly related to the amplitude of the ballistic movement and, therefore, via the resulting photospheric compression, to the intensity of the new main shock produced later around phase 0.881. 
As noted above, this shock is moderate ($V_{\textrm{shock}}=121$\kmsn). Unfortunately, for the previous pulsation cycle, we do not know the shock velocity because the main shock of this cycle occurs about ten hours earlier in the middle of the day. It is to highlight that the Blazhko cycle for these two consecutive pulsations was the same, just after a maximum ($\psi=3.33$). 

\subsubsection{The second H$\alpha$ emission}

\begin{figure}[!ht]
  \centering
  \includegraphics[width=0.97\hsize]{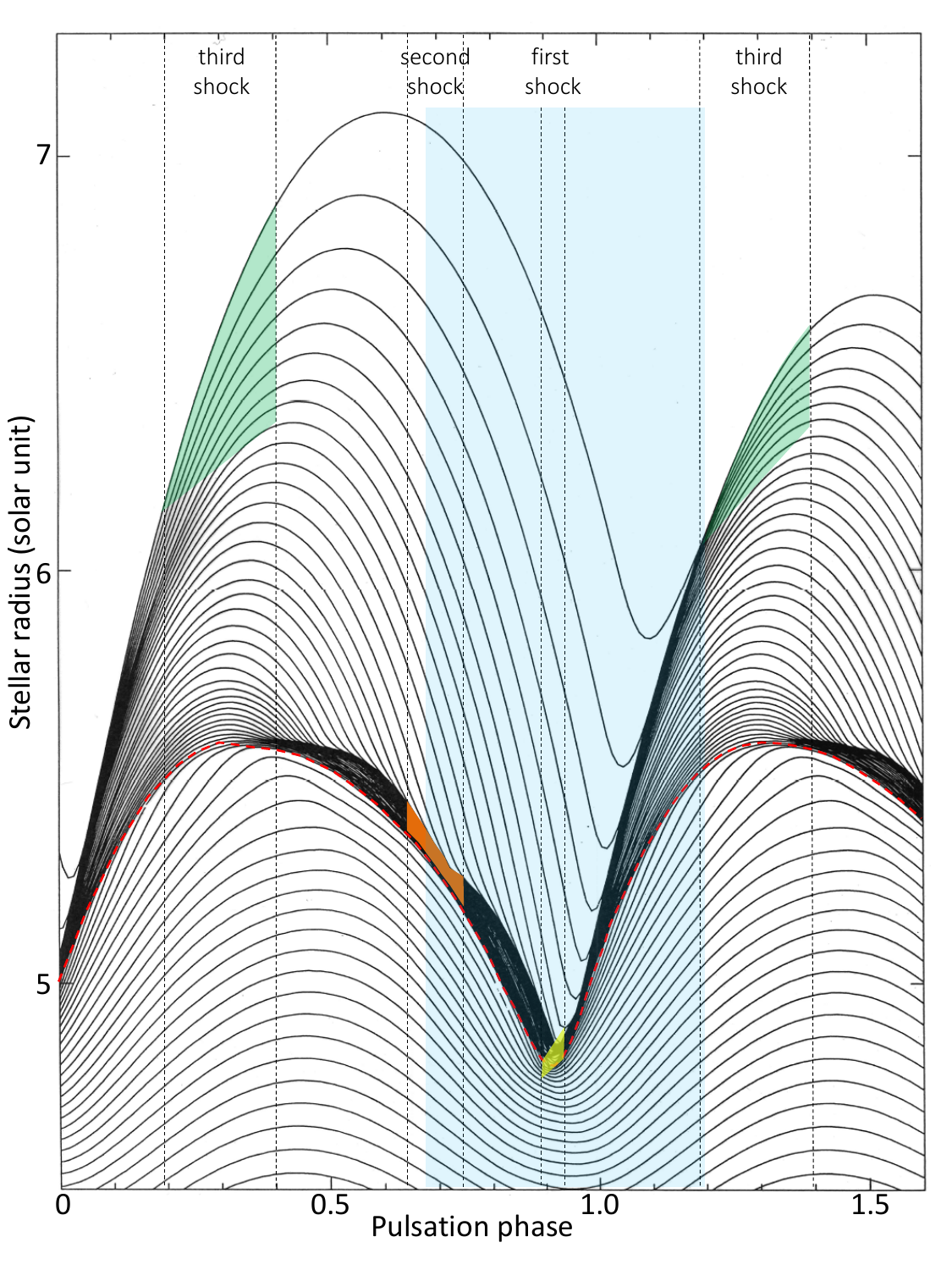}
  \caption{Theoretical radial displacements, in solar units, of the mass zones of the RR Lyr nonlinear nonadiabatic model ''RR41'' \citep{FG1997}. The regions of formation of the three \ha emissions are approximately shown in green, orange, and yellow. The light blue area indicates the limits of the night of August 31, 1997 during which the three \ha emissions were observed (Fig.~\ref{figure9}). The red dashed line approximately represents the location 
of the photosphere defined by the position of the hydrogen ionization front.}
  \label{figure11}
\end{figure}

The blueshifted second emission within the H$\alpha$ profile was first observed by \citet{GC1988} in two stars: X\,Ari (a non-Blazhko RR Lyrae star) and RR\,Lyr itself. It might be the consequence of the colliding upper atmospheric layers within the deeper photospheric ones during the infalling phase of the ballistic motion. This strong atmospheric compression occurs approximately in the pulsation phase interval 0.65-0.75. During this interval, a blueshifted emission is observed, which is called the second emission. During the night of August 31, 1997, its velocity remains approximately constant and equal to 23\kms so weakly supersonic (Mach number of around 2). Consequently, this means that an ascending supersonic shock front occurs within the deep atmosphere (Fig.~\ref{figure11}). Between 1994 and 1997, the second emission was detected seven times (Table \ref{Ha2}). Depending on the night, its velocity varies little but remains constant during the night, between 21 and 29 \kmsn. Although it characterizes the reaction of the atmosphere, the associated Mach number between 2 and 3 is somehow a measure of the compression ratio. The larger it is, the stronger the compression. 

Although there are only five nights with an estimate of the shock velocity (Table \ref{Ha2}), it seems that the lower this velocity, the greater the intensity of the second emission. Of course, this result must be confirmed by other observations. However, if it is correct, it would mean that the larger the compression rate induced by the upper layers of the atmosphere, the greater the density of the photospheric layers. Consequently, the acceleration of the shock induced by the $\kappa$-mechanism would be limited, hence, the velocity of the shock when it leaves the photosphere.

The amplitude of the ballistic movement of the preceding cycle of pulsation would therefore have a notable influence on the intensity of the shock produced during the following cycle. Of course, the intensity of the shock of the initial cycle is directly related to the amplitude of the ballistic movement since the shock contributes to the elevation of the atmospheric layers. Thus, there is a significant hydrodynamic coupling between the pulsation cycles via atmospheric dynamics and the intensity of the shocks -- all the more so as the shocks are strong. A measure of the importance of this coupling is given by the presence and intensity of the second emission.

In the context of the observations in Fig.~\ref{figure9}, the intensity of the second emission depends on the amplitude of the ballistic movement of the “previous” pulsation cycle or, more precisely, on the intensity of the compression that produces this emission. The rate of compression of the photospheric layers induced by the end of the ballistic movement should contribute to the degree of ionization of the subphotospheric layers of hydrogen and helium (He\,I and maybe also He\,II). 
As a result, the intensity of the $\kappa$-mechanism should depend on it, and hence the strength of the resulting new main shock.
Thus, the intensity of the second shock should have an influence on the strength of the main shock which will be produced next. The greater the intensity of the second shock, the more intense the main shock should be. Naturally, the greater the intensity of the main shock, the greater the intensity of the ballistic movement to come should be important, and therefore that of the third shock. Finally, through the interdependence of these three shocks, we may find the principle of an amplification mechanism of the amplitude of the pulsation cycle. 

\begin{figure}[!ht]
  \centering
  \includegraphics[width=0.97\hsize]{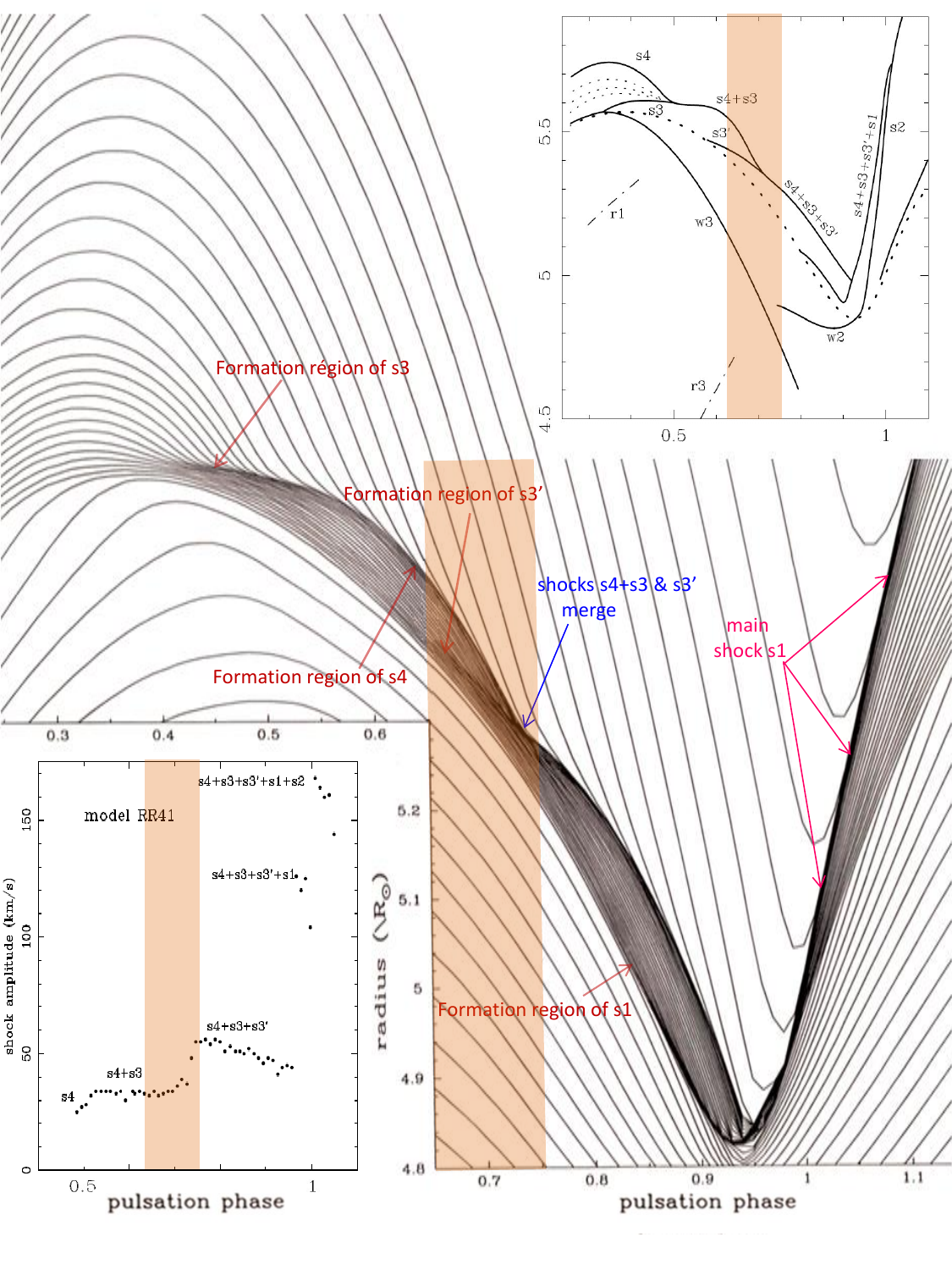}
  \caption{Zoom on Fig.\,11 around the photospheric zone showing the location of the formation region of the different shocks occurring in the atmosphere of RR\,Lyr 
in the framework of the nonlinear nonadiabatic model ''RR41'' \citep{FG1997}. The window of the pulsation phases in which is the second \ha emission is observed is shown by the area colored in light orange. After the phase, at about 0.45 until 0.92, particularly in this area, all shocks identified by the model are receding for the observer.}
  \label{figure12}
\end{figure}

The second \ha emission is blueshifted and always remains blueshifted throughout its observation window, which is typically between $\varphi = 0.65$ and $\varphi = 0.75$. In addition during this phase interval, it maintains an approximately constant velocity between 20 and 30\kms depending on the night. This velocity is weakly supersonic, between two and three times the speed of sound. As shown by \citet{FG1997}, a shock wave propagating in a homogeneous atmosphere with such modest velocity could produce a very weak downstream radiation flux (their Fig.\,11). The shock is not yet a radiative shock for which most of the shock energy is lost by radiation. It is therefore the phenomenon of compression on the dense photospheric layers, which induces a significant heating of the gas. This compression is similar to a collision phenomenon. In order to be able to explain the blueshift of the emission, this compression front must rise in the atmosphere at a supersonic velocity. However, the whole atmosphere and the shocks produced in it fall on the star as demonstrated by the model RR41 from \citet{FG1997}. The dynamics of mass zones (Fig.~\ref{figure12}) does not highlight such a type of ascending compression, which would necessarily be localized just above the photosphere. \citet{FG1997} have calculated more than 20 pulsation models in the parameter range: $M= 0.578M_{\sun}$, $55<L/L_{\sun}<70$, 6875\,K$<T_{eff}<$7205\,K, $X= 0.7$, $Y= 0.299$. None of them clearly exhibited this phenomenon of ascending collision. However, not all the parameters have been explored such as that of the stellar mass, which determines the gravity and therefore the amplitude of the ballistic movement. Indeed, \citet{FG1997} limited their models to a single mass value: $M= 0.578M_{\sun}$. As the existence of the second emission is not in doubt, a complementary theoretical study should be carried out in order to confirm its dynamic origin, while adjusting the value of certain parameters of RR Lyr.

As can be seen in Fig.~\ref{figure12}, a shock denoted s3' is formed approximately when the second emission occurs. Then, just before this emission disappears, this shock merges with s4+s3. As discussed by \citet{FG1997}, the physical origin of s3' is not clear. It is either caused by the 1H-mode or by a running wave r1 arriving from subphotospheric layers. Due to numerical difficulties and the fact that s3 and s3' form almost simultaneously depending on the model, \citet{FG1997} could not study  the physical origin of s3' in greater detail. In any case, s3' has the effect of slowing down the infalling motion of the photospheric layers and, therefore, of accentuating the phenomenon of collision producing the second emission (Fig.~\ref{figure12}). When s3' merges with s4+s3, observations at high spectral and temporal resolutions should perhaps be able to demonstrate an effect on the intensity of the second emission. The quality of our spectra studied in this article is unfortunately insufficient to highlight any disturbance in the second emission when s3' is supposed to occur.

Finally, as it may be noticed in Fig.~\ref{figure12}, the velocity of the shocks s4, s3, and s3' is hardly supersonic, with a range between 10 and 23 \kmsn. Only their cumulative effect gradually increases their velocity when they merge since s4+s3 and s4+s3+s3' reach a maximum value of 35 and 55\kms, respectively. Then s4+s3  approximately maintains the initial velocity constant (35\kmsn) but the velocity of s4+s3+s3' decreases from their maximum initial value (55\kmsn) to about 40\kms when it is abruptly ''absorbed'' by the main shock s1 around $\varphi = 0.92$. The second emission begins to disappear just after the merger of s3' with s4+s3 and just after, the velocity of the new shock is appreciably increased from 35 to 55\kms (Fig.~\ref{figure12}). This increase certainly has the effect of reducing the atmospheric compression ratio and, therefore, that of the intensity of the second emission, only to see it quickly disappear. Just before this fusion of s3', the velocity of the shock s4+s3 remains approximately constant, which has the advantage of making it possible to maintain a sufficient compression ratio to produce the second emission as is observed. It is clear that it would be beneficial to be able to have observations made with a large telescope of the 4-meter class. Indeed, we would then have at the same time a very good S/N, along with optimal spectral and temporal resolution to be able to follow the detailed evolution of the line profile of this second \ha emission, thus making it possible to "solve" the atmospheric dynamics of the pulsating star.

\subsubsection{The third H$\alpha$ emission}

The redshifted third emission within the H$\alpha$ profile was first observed by \citet{Pre2011} in some RR\,Lyrae stars but the cause of it has not yet been identified. \citet{CP2013} detected this redshifted emission both on Blazhko and non-Blazhko RR\,Lyrae stars. These authors hypothesized that a weakly shock wave create this small redshifted emission. It occurs at the beginning of atmospheric compression, during infall. It is a receding shock in Eulerian rest frame. The authors assume that the shock may be created when a hydrogen recombination front fades away near maximum expansion, thus producing an accumulation of several weak compression waves which turn into a shock wave. 
The third emission was detected for the first time on RR\,Lyr on 2013-2014 dataset by \citet{Gillet2017}. The emission intensity is very weak and its profile is a tiny persistent hump in the red side-line profile. From the observations analyzed here (Table \ref{Ha3}), the third emission is observed in a large pulsation phase interval ($\varphi = 1.153$ and $\varphi = 1.453$), depending on the observation conditions. For instance, during the night of August 5, 1994, the third emission was observed from $\varphi = 1.200$ to $\varphi = 1.453, $ namely, \,during a quarter of the pulsation period. This interval practically covers a large part of the expansion phase of the upper layers of the atmosphere. During this phase interval, these layers are far enough away from the photosphere to be detached from it (Fig.~\ref{figure11}). 

During the observation of the third \ha emission on the night of March 26, 2017 \citep{Gillet2017}, the Na\,I $\lambda$5895.92 (D1) absorption line profile was observed double at the pulsation phase 0.227. Consequently, these authors assume that the upper atmosphere is still falling back onto the star at a large supersonic velocity (50\kmsn) and that the shock wave is located between the two Na\,I layers. Unfortunately, as only one spectrum was obtained that night, it is not possible to follow the evolution of the deformation of the two absorption components imposed by the expected Schwarzschild phenomenon (1952). This interpretation was previously offered by \citet{Gillet2019}.

However, \citet{Gillet2019} have shown that this redshifted sodium component is very likely of interstellar origin. Consequently, the explanation suggested by \citet{Gillet2017} is no longer valid: it is not the result of a violent compression and heating of the gas located above the rising shock front. Thus, it appears that the P-Cygni profile of the third H$\alpha$ emission is a ''true'' P-Cygni profile and a consequence of the expansion of an atmospheric shell, induced by the crossing of the shock through the upper layers of the atmosphere. This shell could be likened to the radiative shock wake (but should not be confused with it). 

The redshifted emission of the P-Cygni profile at an average velocity of the order of 40\kms, which remains approximately constant throughout its period of visibility from $\varphi = 1.2$ to $\varphi = 1.4$. This means that the attenuation of the shock energy by geometric dilution and its radiative losses is approximately compensated by the decrease in the atmospheric density as the shock rises in the upper layers. 
It should be noted that the velocity of the expanding shell, which is formed by the shock wake, is weakly supersonic ($M\lesssim 5$). Consequently, in this range of velocities, the radiative losses of the shock are minor and it is especially the geometric dilution that controls the dissipation of the shock energy.

From one night to the next, the maximum shock velocity $V_{3Max}$ changes from 33\kms to 46\kms (Table \ref{Ha3}). This variation does not seem to correlate with the Blazhko phase. This is also true for the emission intensity. Moreover, it is clear that the value of $V_{3Max}$ is not directly correlated with that of $V_{shock}$ (measured around $\varphi=0.92$). A strong shock in the photosphere will therefore not necessarily be intense in the upper atmosphere. This means that the dynamics of the atmospheric layers is very unequal from one pulsation cycle to another, consequently inducing a strong variation in the dissipation of the shock energy. The intensity of the successive shocks occurring in each pulsation cycle is certainly intense enough to generate a fairly chaotic disorder in the ballistic movement of the atmospheric layers. Consequently, the velocity field in the atmosphere must become highly turbulent for a large fraction of the pulsation period during and immediately after the shock has passed through the atmosphere.

Depending of the pulsation cycle, the shock velocity, $V_{shock}$, is between 100 and 150 \kms when the shock emerges from the photospheric layers around $\varphi = 0.90$. Then it decreases to 50 \kms at $\varphi = 1.04$ and finally finishes at 40 \kms between $\varphi = 1.20$ and $\varphi = 1.40$. During this phase interval between 0.90 and 1.40, the atmosphere is expanding and the shock travels a distance around 40\% of the photospheric radius. This distance is consistent with that given by the RR41 model (Fig.~\ref{figure11}). This result is also consistent with the supposed true nature of the P-Cygni profile of the third emission since, at such a distance from the photosphere, the shock is largely detached from the stellar disk.

\section{Conclusion}

The principal aim of this paper is to use from spectral observations to determine  the connection of the three \ha emission line components occurring in the atmosphere of \rrly during its pulsation cycle with atmospheric dynamics. 

Three main shocks per pulsation period have been observed. The strongest one, called  the "main shock," appears near phase $\varphi=0.9$ and is the consequence of the mechanisms which are at the origin of the pulsation. The second one, called "early shock," has been explained by the strong compression between the supersonic free-infalling motion of the highest atmospheric layers on the slower photospheric layers around phase $\varphi=0.7$. The physical origin of these two shocks was previously identified. 
However, it is the third emission, occurring around phase $\varphi=0.3$, that is the subject of the reanalysis carried out in this study. It is caused by the movement of the upper layers of the atmosphere which are drained by the passage of the main shock wave. With these layers being detached from the photosphere, a P-Cygni profile is thus visible in the spectrum for about 20\% of the pulsation period. The layers contributing to the emission constitute an expanding shell, which could be identified as the radiative wake of the shock. We also find that the supersonic compression (secondary shock) induced the second \ha emission 
occurs at an approximately constant velocity (Mach number between 2 and 3).

The velocity of the shock front was estimated between 100 and 150 $\pm 10$\kms,\ namely, for a Mach number between 10 and 15. It has been established that the shock velocity increases from the
minimum Blazhko to its maximum and afterward, it gradually decreases to the Blazhko minimum to start growing again. However this result should be confirmed using data with better S/N while retaining the best possible resolving power. This observational result would be consistent with the shock model proposed by \citet{Gil2013} to explain the Blazhko effect.

It has been observed that over the same Blazhko cycle and from its minimum, the intensity of the \ha emission and D3 helium emission decrease with the Blazhko phase. This decrease occurs in spite of the shock velocity continuously increasing. On the other hand, the intensity of these two lines begins with the increase in the shock velocity, then reaching a maximum value when $V_s\simeq 137$\kms, and finally decreasing to the maximum Blazhko. Thus, it appears that the intensity of emission lines are modulated by Blazhko effect. This latter phenomenon is probably the physical consequence of the increase in the ionization rate of hydrogen and helium when the shock velocity increases. 

Finally, with the spectroscopic survey presented in this paper, we have elucidated the physical origin of the second and third \ha emissions. Moreover, we have shown that the evolution of the \ha line profiles during the Blazhko cycle is in good agreement with the explanation of the Blazhko effect proposed by \citet{Gil2013}, namely, the shock model. 

In conclusion, the study presented in this article clearly demonstrates that observations at high temporal and spectral resolutions on larger telescopes with a higher S/N 
are therefore essential in attaining a comprehensive understanding the details of the atmospheric dynamics of RR\,Lyrae stars, and of RR\,Lyr in particular. Similarly, the development of more 
detailed nonlinear stellar pulsation models including the generation and propagation of hypersonic shock waves would be essential to confirm the determining role 
of the first overtone when the Blazhko effect is generated.

\bigskip
\begin{acknowledgements}
We thank the Facult\'e des Sciences Semlalia, D\'epartement de Physique, LPHEA, Marrakech and the French OHP-CNRS/PYTHEAS for their support.
The present study has used the SIMBAD database operated at the Centre de Donn\'ees Astronomiques (Strasbourg, France) and 
the GEOS RR\,Lyr database hosted by IRAP (OMP-UPS, Toulouse, France), created by J.F. Le Borgne.
We gratefully acknowledge Christian Feghali for his careful reading of the first version of the manuscript and for his pertinent remarks and Jean-S\'ebastien Devaux for his help.
This work was carried out in part within the framework of the \textit{The \textbf{G}roupe de \textbf{R}echerche sur \textbf{RR} Lyrae} (GRRR) which is an association of professionals and amateur astronomers leading high-resolution spectroscopic and photometric monitoring of complex phenomena such as the \rrlyr Blazhko effect.

\end{acknowledgements}


\bibliographystyle{aa}
\bibliography{rr_lyr_bibliography}

\end{document}